\documentclass[twocolumn,english,aps,prb,twocolum,superscriptaddress,nobibnotes,amsmath,amssymb,floatfix]{revtex4-1}
\usepackage[colorlinks=true,citecolor=blue,linkcolor=magenta]{hyperref}

\usepackage[markup=nocolor, authormarkupposition=left]{changes}
\usepackage{soul}
\usepackage[utf8]{inputenc}
\usepackage[english]{babel}
\usepackage{amsmath,amsfonts,amssymb}
\usepackage[T1]{fontenc}
\usepackage{url}

\usepackage{amsmath}
\usepackage{amsfonts}
\usepackage{amssymb}
\usepackage{upgreek}

\usepackage{epstopdf}
\usepackage{graphicx}
\graphicspath{{./Figures/}}

\begin{document}
\title{Magnetic-Free Silicon Nitride Integrated Optical Isolator}

\author{Hao Tian}
\thanks{These authors contributed equally to this work.}
\affiliation{OxideMEMS Lab, Purdue University, 47907 West Lafayette, IN, USA}

\author{Junqiu Liu}
\thanks{These authors contributed equally to this work.}
\affiliation{Institute of Physics, Swiss Federal Institute of Technology Lausanne (EPFL), CH-1015 Lausanne, Switzerland}

\author{Anat Siddharth}
\affiliation{Institute of Physics, Swiss Federal Institute of Technology Lausanne (EPFL), CH-1015 Lausanne, Switzerland}

\author{Rui Ning Wang}
\affiliation{Institute of Physics, Swiss Federal Institute of Technology Lausanne (EPFL), CH-1015 Lausanne, Switzerland}

\author{Terence Blésin}
\affiliation{Institute of Physics, Swiss Federal Institute of Technology Lausanne (EPFL), CH-1015 Lausanne, Switzerland}

\author{Jijun He}
\affiliation{Institute of Physics, Swiss Federal Institute of Technology Lausanne (EPFL), CH-1015 Lausanne, Switzerland}

\author{Tobias J. Kippenberg}
\email[]{tobias.kippenberg@epfl.ch}
\affiliation{Institute of Physics, Swiss Federal Institute of Technology Lausanne (EPFL), CH-1015 Lausanne, Switzerland}

\author{Sunil A. Bhave}
\email[]{bhave@purdue.edu}
\affiliation{OxideMEMS Lab, Purdue University, 47907 West Lafayette, IN, USA}

\maketitle

\noindent\textbf{Integrated photonics \cite{Thomson:16} has enabled signal synthesis, modulation and conversion using photonic integrated circuits (PIC). 
Many materials have been developed, among which silicon nitride (Si$_3$N$_4$) has emerged as a leading platform particularly for nonlinear photonics \cite{Moss:13, Gaeta:19}. 
Low-loss Si$_3$N$_4$ PIC has been widely used for frequency comb generation \cite{Kippenberg:18, Gaeta:19}, narrow-linewidth lasers \cite{Gundavarapu:19, Jin:21, Xiang:21}, microwave photonics \cite{Roeloffzen:13, Wu:18}, photonic computing networks \cite{Feldmann:21, Arrazola:21}, and even surface-electrode ion traps \cite{Mehta:20, Niffenegger:20}. 
Yet, among all demonstrated functionalities for Si$_3$N$_4$ integrated photonics, optical non-reciprocal devices, such as isolators and circulators, have not been achieved. 
Conventionally, they are realized based on Faraday effect of magneto-optic materials \cite{Srinivasan:18, Bi:11, Huang:16, Yan:20} under external magnetic field. 
However, it has been challenging to integrate magneto-optic materials that are not CMOS-compatible and that require bulky external magnet. 
Here, we demonstrate a magnetic-free optical isolator based on aluminum nitride (AlN) piezoelectric modulators monolithically integrated on ultralow-loss Si$_3$N$_4$ PIC \cite{Liu:20b}. 
The transmission reciprocity is broken by spatio-temporal modulation \cite {Yu:09} of a Si$_3$N$_4$ microring resonator with three AlN bulk acoustic wave resonators \cite{Tian:20, Liu:20a} that are driven with a rotational phase. 
This design creates an effective rotating acoustic wave that allows indirect interband transition \cite {Kang:11, Kittlaus:18, Sohn:18} in only one direction among a pair of strongly coupled optical modes. 
Maximum of 10 dB isolation is achieved under 100 mW RF power applied to each actuator, with minimum insertion loss of 0.1 dB. 
The isolation remains constant over nearly 30 dB dynamic range of optical input power, showing excellent optical linearity.
Our integrated, linear, magnetic-free, electrically driven optical isolator could become key building blocks for integrated lasers \cite{HuangD:19, Xiang:20, McKinzie:21, Xiang:21, Liang:21}, chip-scale LiDAR engines \cite{Riemensberger:20, Lukashchuk:21}, as well as optical interfaces for superconducting circuits \cite{Awschalom:21}. 
}

Integrated photonics \cite{Thomson:16} has enabled chip-scale optical systems with compact sizes, portable weight, and low power consumption, which have been translated from table-top research setups to commercial products. 
With the advancement of nano-fabrication technology, there have been major achievements in integrated semiconductor lasers \cite{HuangD:19, Xiang:20, McKinzie:21, Xiang:21, Liang:21}, modulators \cite{Melikyan:14, WangC:18, HeM:19, Tian:20, Chen:14, Tadesse:14, Shao:19, Tian:21}, photodetectors\cite{Kang:09, Beling:16}, as well as ultralow-loss photonic integrated circuits (PIC) \cite{Zhang:17, Yang:18, Chang:20, Xuan:16, Ji:17, Liu:20b, Ye:19b, Puckett:21}. 
Importantly, the nonlinearities of ultralow-loss PICs have been explored and harnessed, giving rise to nonlinear photonics such as optical frequency comb (OFC) generation \cite{Moss:13, Kippenberg:18, Gaeta:19}.  
Formed in Kerr-nonlinear optical microresonators, dissipative Kerr solitons (DKS) microcombs \cite{Kippenberg:18, Gaeta:19, Herr:14} constitute miniaturized, coherent, broadband OFCs with repetition rates in the millimeter-wave to microwave domain, and are amenable to heterogeneous or hybrid integration with III-V/Si lasers\cite{Stern:18, Raja:19, Shen:20, Xiang:21}. 
Major progress has been made in the past decade in developing various platforms \cite{Kovach:20}, ranging from Si$_3$N$_4$\cite{Xuan:16, Ji:17, Liu:20b, Ye:19b}, LiNbO$_3$\cite{Wang:19, Zhang:19, He:19, Gong:20}, AlN\cite{Jung:13, LiuX:18} and AlGaAs\cite{Pu:16, Chang:20}. 
Among the materials developed thus far for integrated nonlinear photonics, Si$_3$N$_4$ has become the leading platform due to its exceptionally low loss (down to 1 dB/m), high Kerr nonlinearity, absence of two-photon absorption, and high power handling capability. 
Indeed, most of the system-level demonstrations of microcombs have been based on Si$_3$N$_4$, such as coherent communication~\cite{Marin-Palomo:17, Fulop:18, Corcoran:20}, astronomical spectrometer calibration~\cite{Obrzud:19, Suh:19}, distance measurement \cite{Trocha:18, Suh:18, Jang:21, Wang:20}, coherent LiDAR \cite{Riemensberger:20, Lukashchuk:21}, frequency synthesizers~\cite{Spencer:18, Liu:20}, optical atomic clocks \cite{Newman:19}, and photonic neromorphic computing \cite{Feldmann:21, Xu:21}. 
Meanwhile, high-$Q$ Si$_3$N$_4$ microresonators have also been used recently to create ultralow-noise semiconductor lasers \cite{Jin:21, Xiang:21} whose performance is on par with advanced fiber lasers.

\begin{figure*}[ht]
\centering
\includegraphics{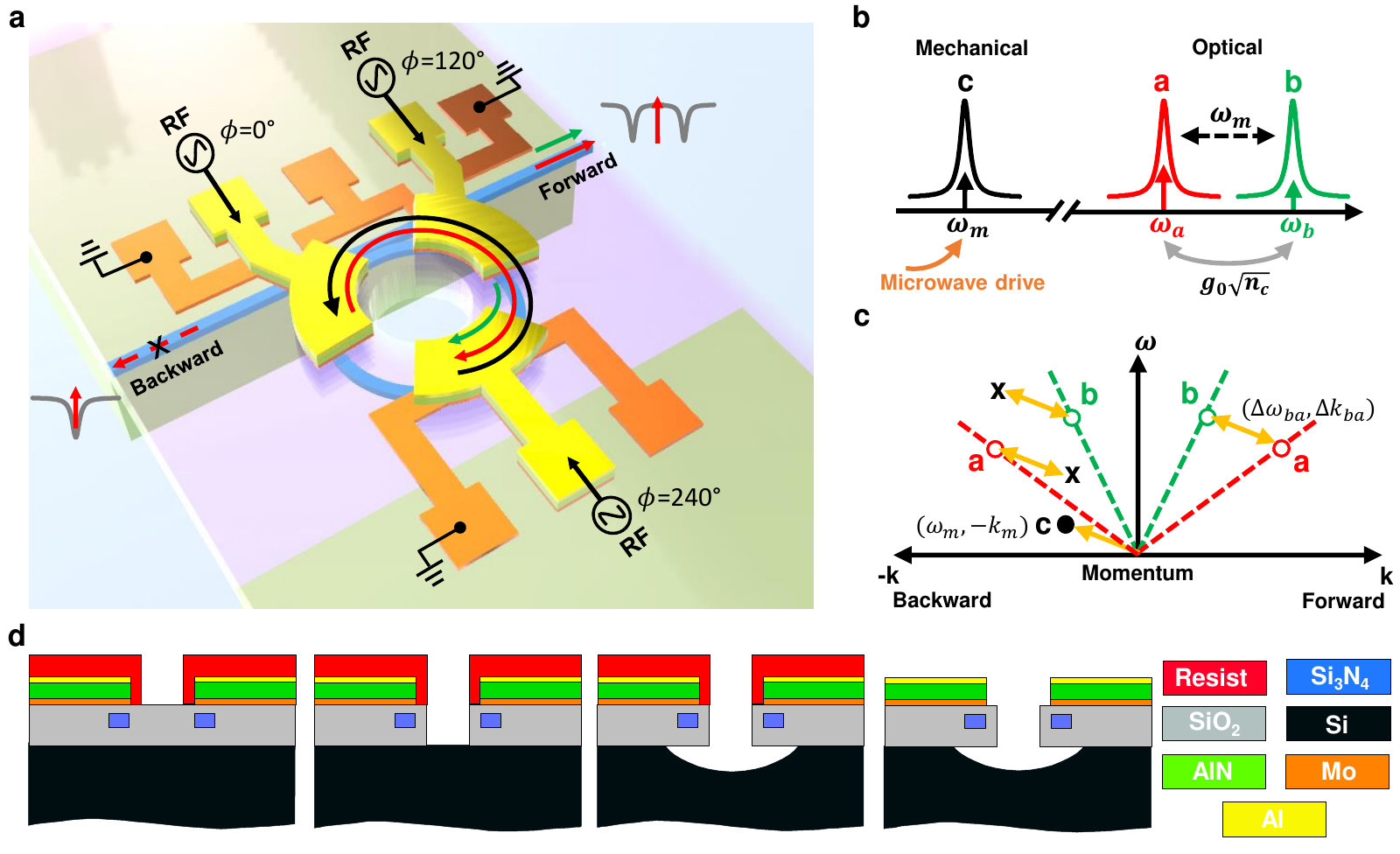}
\caption{
\textbf{Principle of the nitride optical isolator}. 
\textbf{a} Schematic and device rendering. 
Three discrete AlN piezoelectric actuators are equidistantly placed on top of an integrated Si$_3$N$_4$ microring resonator (blue solid).
Upon coherent driving these actuators with fixed relative phases $(\phi_1,\phi_2,\phi_3)=(0^\circ, 120^\circ, 240^\circ)$, a rotating acoustic wave (black arrow) is generated that spatial-temporal-modulates the two co-propagating optical modes (red and green arrows), leading to a indirect interband transition in only one direction. 
\textbf{b} Frequency domain representation illustrating the indirect interband transition. When the two optical modes, $a$ and $b$, are spaced by the resonant frequency $\omega _m$ of the mechanical mode $c$, i.e. $\Delta \omega_{ba}=\omega_b-\omega_a=\omega _m$, scattering among modes $a$ and $b$ happens with a scattering rate of $g=g_0\sqrt{n_c}$ under a microwave drive at $\omega_m$. 
\textbf{c} Schematic of $\omega -k $ space showing the phase matching condition, i.e. the energy ($\Delta \omega _{ba}=\omega _m$) and momentum ($\Delta k _{ba}=-k _m$) conservations. 
Interband transition that couples two optical modes with the acoustic wave is only allowed in the direction where phase matching condition is fulfilled, giving rise to transparency on resonance in one direction (``forward'', as shown in \textbf{a}) and extinction in the other direction (``backward''). 
\textbf{d} Fabrication process flow to suspend the SiO$_2$ membrane containing the Si$_3$N$_4$ microring resonator, to enhance the optomechanical coupling.
The key technique used here is Si isotropic dry etch using SF$_6$ Bosch process. 
}
\label{Fig:1}
\end{figure*}

Despite of these advances of Si$_3$N$_4$ integrated photonics, non-reciprocal devices, such as isolators and circulators that are widely used in optical communications and datacenters for signal routing, multiplexing, and protecting lasers from reflections, have not been improved via these developments.
Conventionally, optical non-reciprocity is realized in magneto-optic materials \cite{Srinivasan:18} where the Faraday effect induces non-reciprocal polarization rotation under an external magnetic field. 
However, magneto-optic materials are not CMOS-compatible, posing challenges to integrate using standard CMOS techniques developed for silicon photonics. 
Besides, the magneto-optic effect is weak from the near-infrared to visible wavelength range, therefore requiring a strong external magnetic field applied at microscopic scale. 
Nevertheless, successful integration of cerium-substituted yttrium iron garnet (Ce:YIG) on PIC (e.g., on Si \cite{Bi:11, Huang:16} or Si$_3$N$_4$ \cite{Yan:20}) has been demonstrated via wafer bonding or deposition, though these efforts suffer from significant optical losses and external magnet. 
In addition, the requirement of large magnetic field bias makes it incompatible with superconducting qubits where optical isolation is needed for blocking reflected noises in optical interfaces for quantum interconnects \cite{Awschalom:21}. 

To overcome these bottlenecks, magnetic-free schemes have been demonstrated to break Lorentz reciprocity or time reversal symmetry, through synthetic magnetic field \cite{Fang:12, Tzuang:14, Kim:21, Dostart:21}, optical nonlinearities \cite{Grigoriev:11, Chang:14, DelBino:18, Hua:16, YangKY:20, Cao:20, Shao:20}, optomechanically induced transparency \cite{Hafezi:12, Shen:16, Ruesink:16, Shen:18, Ruesink:18}, and stimulated Brillouin scattering (SBS) \cite{Kang:11, Dong:15, Kim:15, Kim:17}. 
In addition, optical non-reciprocity has also been demonstrated in atomic systems \cite{Scheucher:16, ZhangS:18, HuX:20}. 
However, challenges remain in all these approaches. 
For example, optomechanically induced non-reciprocity requires air-cladded, isolated microtoroids or microspheres that are difficult to integrate with PICs.  
In addition, the isolation bandwidth is primarily limited by the mechanical resonance linewidth that is typically within kilohertz to megahertz range. 
Separating the signal from the control pump laser using common multiplexing schemes is a formidable challenge, as they are spaced by the mechanical resonance that is below hundred megahertz.  
While nonlinear optics can work passively without active modulation, the main concern is the dynamic reciprocity that forbids the propagation of light in both directions with limited dynamic range of input optical power \cite{Shi:15}. 


Spatio-temporal modulation \cite{Yu:09, Lira:12, Sounas:17, Sounas:14, Shi:18, Kittlaus:18, Sohn:18, Kittlaus:21}, which breaks the reciprocity by coupling two optical modes and prescribing phase matching condition by the active modulation, stands out in terms of integration and applicability on nearly all optical materials. 
Recently, schemes based on acousto-optic modulation (AOM) has been extensively developed due to the compatibility of this approach with low-loss PICs (e.g., AlN \cite{Sohn:18} and Si \cite{Kittlaus:21}). 
Thus far, only non-reciprocal sideband modulation is achieved, limited by the modulation efficiency and power handling capability of the interdigital transducers (IDT) for generating surface acoustic waves (SAW).

In this work, we demonstrate the first AOM-based optical isolator for Si$_3$N$_4$ integrated photonics.
Three AlN piezoelectric actuators are equidistantly placed along a Si$_3$N$_4$ microring resonator, and generate high-overtone bulk acoustic resonances (HBAR) to realize AOM \cite{Tian:20, Liu:20a, Tian:21}. 
By carefully tuning the relative drive phases among these actuators, HBAR modes create an effective rotating acoustic wave that couples two optical modes in the momentum-biased direction. 
We achieve an optical cooperativity $C=1$ with 14 dBm RF drive power applied on each actuator, and maximum isolation ratio of 10 dB and minimum insertion loss of 0.1 dB with 20 dBm RF power. 
An isolation bandwidth of 700 MHz is obtained, which is primarily determined by the optical resonance linewidth. 
The device performance can be further improved by designing and operating the Si$_3$N$_4$ microresonator in the critical coupling regime.

\section{Results}

\noindent\textbf{Device principle and design}. 
Our integrated optical isolator consists of three AlN piezoelectric actuators on top of a Si$_3$N$_4$ microring resonator, as shown in Fig. \ref{Fig:1}a. 
A Si$_3$N$_4$ PIC was fabricated using the photonic Damascene process \cite{Pfeiffer:18b, Liu:18a, Liu:20b}, followed by monolithic integration of AlN actuators as illustrated in Ref. \cite{Tian:20, Liu:20a}. 
The piezoelectric AlN thin film (green), with a piezoelectric coefficient \cite{Dubois:99} of $d_{33}=3.9$ pm V$^{-1}$, is sandwitched between top aluminum (Al, yellow) and bottom molybdenum (Mo, orange) electrodes. 
When the electrodes are microwave-driven, bulk acoustic waves are formed vertically in the substrate (i.e. the HBAR mode) beneath the actuators.
The Si$_3$N$_4$ PIC (blue) is fully buried in thick SiO$_2$ cladding to preserve the optical losses. 
In contrast to the previous work \cite{Tian:20, Liu:20a}, here we apply a Si-release process to create suspended SiO$_2$ membrane where Si$_3$N$_4$ microresonator is embedded. 
The process flow is shown in Fig. \ref{Fig:1}d. 
This release process enables tight confinement of HBAR modes inside the Fabry-P$\acute{\text{e}}$rot acoustic cavity formed by the top and bottom SiO$_2$ surfaces, and thus enhances the acousto-optic coupling through stress-optical effect, as shown in Ref. \cite{Tian:20, Tian:21}. 

The Si$_3$N$_4$ microresonator is designed to support two optical eigenmodes, $a$ and $b$, with frequency difference matching a mechanical/acoustic resonant frequency, as shown in Fig. \ref{Fig:1}b. 
The microwave drive applied on the three AlN actuators creates acoustic waves inside the mechanical cavity, which scatters light between modes $a$ and $b$ (i.e. indirect interband transition). 
Figure \ref{Fig:1}c illustrates the $\omega -k $ space, where $k=2\pi /\lambda$ is wavenumber (the photon/phonon momentum is $\hbar k$, with the sign denoting the rotating direction along the microring, clockwise or counter-clockwise), and $\omega$ is the angular frequency. 
To induce interband transition, energy and momentum conservations must be satisfied, known as the ``phase matching condition''. 
With a nonzero phonon momentum $k_m$, phase matching requires $\Delta \omega _{ba}=\omega _b - \omega _a =\omega _m$ and $\Delta  k _{ba}= k _b - k _a = -k _m$, where the minus sign of $k_m$ indicates that the acoustic wave needs to counter-propagate with the two co-propagating optical modes, as illustrated in Fig. \ref{Fig:1}a. 
We denote this direction, where phase matching condition is fulfilled, as the ``forward'' direction in the following discussion. 

While a single vertical HBAR mode from one AlN actuator carries zero in-plane momentum, an effective acoustic wave rotating along the microring is generated by driving three actuators coherently with phases of ($0^\circ$, $120^\circ$, $240 ^\circ$), i.e. $120^\circ$ phase difference between two adjacent actuators, as illustrated in Fig. \ref{Fig:1}a. 
In the forward direction, two optical modes are coupled via the acoustic wave, which can be described by a simplified Coupled Mode Theory (CMT) equations \cite{Fan:18}:
\begin{eqnarray}
\label{eq1}
\dot{a} & = & -(i\Delta _a +\frac{\kappa _{a}}{2})a-igbe^{i\omega _d t} +\sqrt{\kappa _{a,\text{ex}}}a_\text{in}\\
\label{eq2}
\dot{b} & = & -(i\Delta _b +\frac{\kappa _{b}}{2})b-igae^{-i\omega _d t}
\end{eqnarray}
where $a$ ($b$) and $\kappa _{a}$ ($\kappa _{b}$) are the intra-cavity amplitude and the total loss rate of mode $a$ ($b$), $\kappa _{a,\text{ex}}$ is the external coupling rate of mode $a$. 
These equations of motion are transformed under the rotating wave approximation (RWA) around the input laser frequency $\omega _L/2\pi$, thus $\Delta _a=\omega _a -\omega _L$ ($\Delta _b=\omega _b -\omega _L$) is the laser frequency detuning to the resonant frequency of $a$ ($b$). 
$\omega _d/2\pi$ is the microwave drive frequency which can be slightly detuned from the mechanical resonant frequency $\omega _m/2\pi$. 
Mode $a$ is probed by the input light with amplitude $a_\text{in}$, and its optical transmission/isolation is studied in the following analysis.

In the forward direction, the two optical modes undergo interband transition with a scattering rate $g=g_0 \sqrt{n_c}$, where $g_0$ is the single phonon-photon coupling rate describing the optomechanical interaction strength, and $n_c$ is the intra-cavity phonon number. 
The optical cooperativity, i.e. the ratio of scattering rate and losses, is $C=4g^2/\kappa _a \kappa _b$. 
Strong coupling requires $C \gg 1$, and can leads to Rabi oscillation and mode splitting.
The latter results in a transparency window on the resonance in the light transmission, which can be calculated from Eqs. \ref{eq1} and \ref{eq2} as follows:
\begin{equation}
    T|_{\Delta _a=0}=\left| \frac{a_\text{out}}{a_\text{in}} \right| ^2=\left[ 1-\frac{2\kappa _{a,\text{ex}}}{\kappa _a(1+C)}\right] ^2
\label{eq3}
\end{equation}
In the forward direction, $C \gg 1$ and $T|_{\Delta _a=0}=1$. 
This transparency can be understood intuitively as the impedance mismatch between the bus waveguide and the microresonator, which is resulted from the increasing effective intrinsic loss due to the scattering to the other optical mode. 
In the backward direction where the three-wave phase matching is not fulfilled, interband transition is prohibited, leading to $C \ll 1$ and $T|_{\Delta _a=0} =0$ in the critical coupling regime ($\kappa _a=2\kappa _{a,\text{ex}}$). 
Consequently, the microresonator remains critically coupled and its light transmission is not affected by the presence of the acoustic wave. 
This non-reciprocal transmission between the forward ($T=1$) and backward directions ($T=0$) is the basic of our optical isolator.

\begin{figure*}[t!]
\centering
\includegraphics{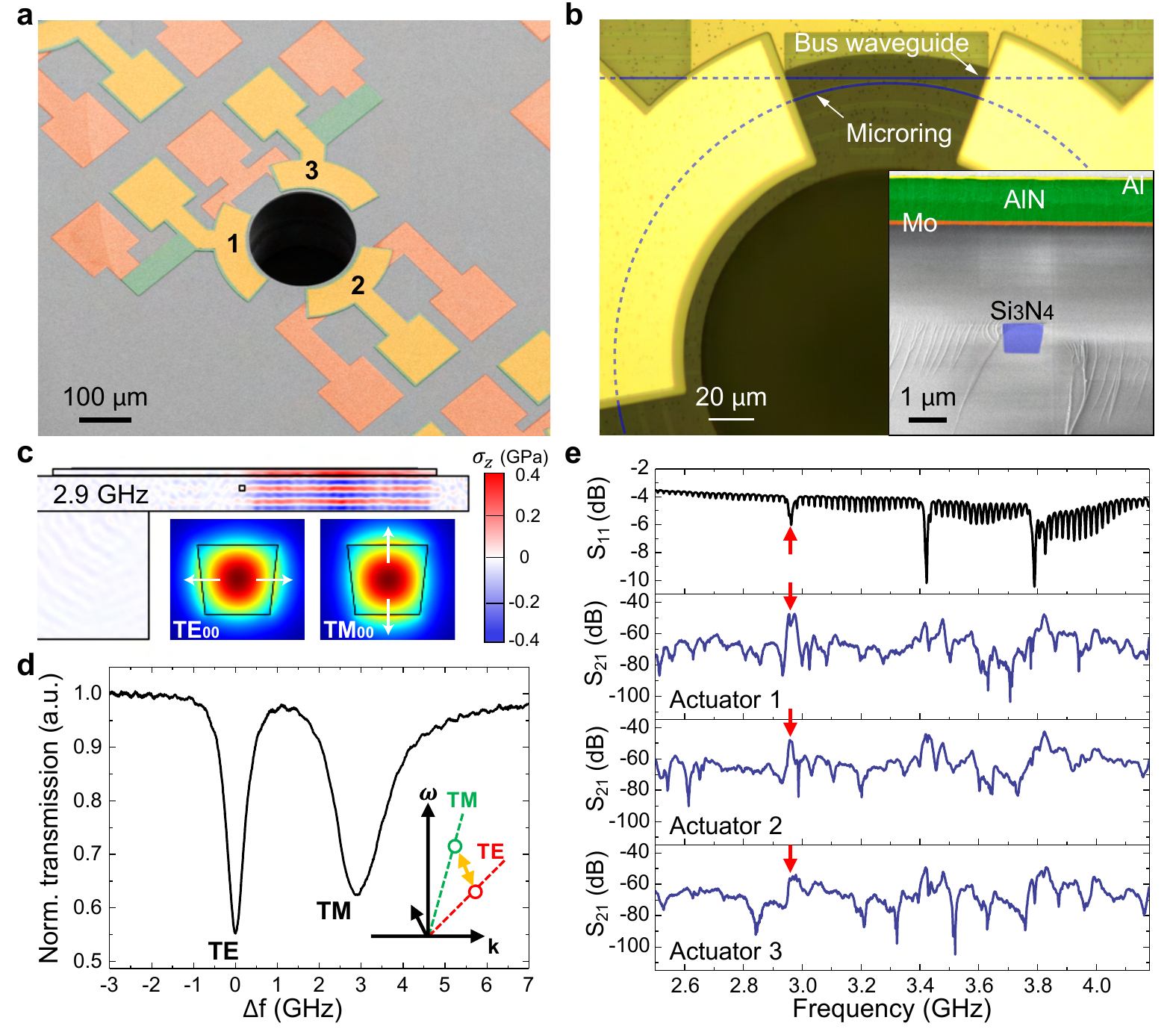}
\caption{
\textbf{Characterization of optical and mechanical properties of the isolator device}. 
\textbf{a} False-colored top-view SEM image of the fabricated device. 
\textbf{b} Optical microscope image highlighting bus-microring coupling section, the released SiO$_2$ area, and the relative positions between Si$_3$N$_4$ waveguides (blue line) and two actuators. 
Inset: False-colored SEM image of the sample cross-section, showing the vertical structure of the piezoelectric actuator and quasi-square Si$_3$N$_4$ photonic waveguide.
\textbf{c} FEM numerical simulations of the vertical stress $\sigma _z$ distribution of a typical HBAR mode at 2.9 GHz, and the optical profiles of the TE$_{00}$ and TM$_{00}$ modes of the quasi-square Si$_3$N$_4$ waveguide. 
White arrows mark the optical polarization directions.
\textbf{d} Optical transmission spectrum showing a pair of TE$_{00}$ and TM$_{00}$ modes with frequency spacing of around 3 GHz. 
The x-axis is frequency-calibrated relative to the center frequency of TE$_{00}$ mode around 1546 nm. 
Inset shows the schematic of relative position of the two modes in $\omega -k$ space. 
\textbf{e} From top to bottom are microwave reflection S$_{11}$, and optomechanical response S$_{21}$ of the actuator 1, 2, 3 (labeled in \textbf{a}), respectively. 
Red arrows mark the mechanical mode at 2.958 GHz that is used in the experiment. 
}
\label{Fig:2}
\end{figure*}

Meanwhile, in the forward direction, a single modulation sideband is generated in mode $b$ which is frequency-shifted by $\omega _d$ relative to the input laser $\omega _L/2\pi$. 
The mode conversion efficiency $\eta$ is:
\begin{equation}
    \eta |_{\Delta _a=0}= \left| \frac{b_\text{out}}{a_\text{in}} \right| ^2 = \frac{\kappa _{a,\text{ex}}}{\kappa _a}\frac{\kappa _{b,\text{ex}}}{\kappa _b}\frac{4C}{(1+C)^2}
\label{eq4}
\end{equation}
Equations \ref{eq3} and \ref{eq4} are derived assuming that the microwave drive $\omega _d$ matches the optical frequency spacing $\Delta \omega _{ba}$, i.e. $\omega _d=\Delta \omega _{ba}$. 
A detailed derivation of these equations and general cases with a frequency mismatch, which were used for subsequent fitting experiments, are provided in the Methods section. 
This non-reciprocal sideband generation has been demonstrated in previous studies in the $C < 1$ regime \cite{Kittlaus:18, Sohn:18, Kittlaus:21}.

\noindent \textbf{Experimental realization}. 
Figure \ref{Fig:2}a shows the false-colored, top-view scanning electron microscope (SEM) image of the fabricated device with three AlN actuators integrated on a released Si$_3$N$_4$ microring resonator. 
The thickness of Al / AlN / Mo is 100 / 1000 / 100 nm, respectively. 
The center black hole is opened for Si-isotropic dry etching using the SF$_6$ Bosch process, to partially remove the Si substrate and to suspend the 5.4-$\upmu$m-thick SiO$_2$ cladding. 
Figure \ref{Fig:2}b shows the optical microscope image highlighting the bus waveguide coupling region, the Si$_3$N$_4$ microring with a radius of 118 $\upmu$m buried in the suspended SiO$_2$ membrane, and two AlN actuators. 
Figure \ref{Fig:2}c shows the simulated stress distribution of one HBAR mode within the SiO$_2$ layer using Finite Element Method (FEM). 
It can be seen that the HBAR mode is uniformly distributed under the AlN actuator and tightly confined in the SiO$_2$ membrane, allowing direct modulation of the optical mode propagating along the waveguide through photoelastic and moving boundary effects \cite{Balram:14, Jin:18, Stanfield:19}. 

We use the fundamental transverse electric (TE$_{00}$, mode $a$) and magnetic modes (TM$_{00}$, mode $b$) to realize interband transition assisted with the acoustic wave.  
A quasi-square waveguide cross-section ($810 \times 820$ nm$^2$) is used, as shown in Fig. \ref{Fig:2}b inset. 
Figure \ref{Fig:2}c shows the simulated TE$_{00}$ and TM$_{00}$ mode profiles, which also include the slanted sidewall. 
The transmission of polarization-tilted light through the optical microresonator, including a pair of TE$_{00}$ and TM$_{00}$ resonances, is shown in Fig. \ref{Fig:2}d. 
The TM$_{00}$ mode frequency is 3 GHz higher than that of the TE$_{00}$ mode at around 1546 nm wavelength. 
The resonance linewidth is 1.16 (0.68) GHz for the TM$_{00}$ (TE$_{00}$) mode. 

Figure \ref{Fig:2}e shows the microwave reflection S$_{11}$, where mechanical resonances are revealed. 
Only one actuator's S$_{11}$ is shown as the others are similar. 
Three strong resonances are found around 3.0, 3.4 and 3.8 GHz, which are due to the SiO$_2$ mechanical cavity with $\sim$470 MHz free spectral range (FSR, determined by the SiO$_2$ cladding thickness). 
Besides, weak resonances with an FSR of $\sim$19 MHz are observed, due to the HBARs in thick Si substrate formed under square signal probe pads which were not undercut \cite{Tian:20}. 
However, only the HBARs confined in SiO$_2$ can efficiently modulate the optical mode because the HBARs in the Si substrate have negligible overlap with the Si$_3$N$_4$ waveguide, which can be verified from the optomechanical S$_{21}$ response shown in Fig. \ref{Fig:2}e. 
S$_{21}$ measures the ratio between the output light intensity modulation and the microwave drive power. 
Three actuators are measured individually. 
Since the HBARs are mainly determined by the thickness of each layer that was well controlled during fabrication and highly uniform over the full wafer scale, the HBAR frequencies of the three actuators show only sub-megahertz misalignment. 
A maximum of -45 dB S21 was achieved, providing an improvement of 20 dB over a previously reported unreleased Si HBAR AOM \cite{Tian:20}. 
This is due to the significantly reduced mechanical mode volume and tighter HBAR confinement in the released SiO$_2$ membrane. 
The SiO$_2$ HBAR at 2.958 GHz is used in the following experiments to match the optical mode spacing. 
Additionally, the signal cross-talk between the actuators is maintained below -60 dB (see Supplementary Note 6) because the HBARs are tightly confined vertically beneath the actuator, and the center etched hole prevents the transmission of any lateral mechanical modes. 

\begin{figure*}[t!]
\centering
\includegraphics{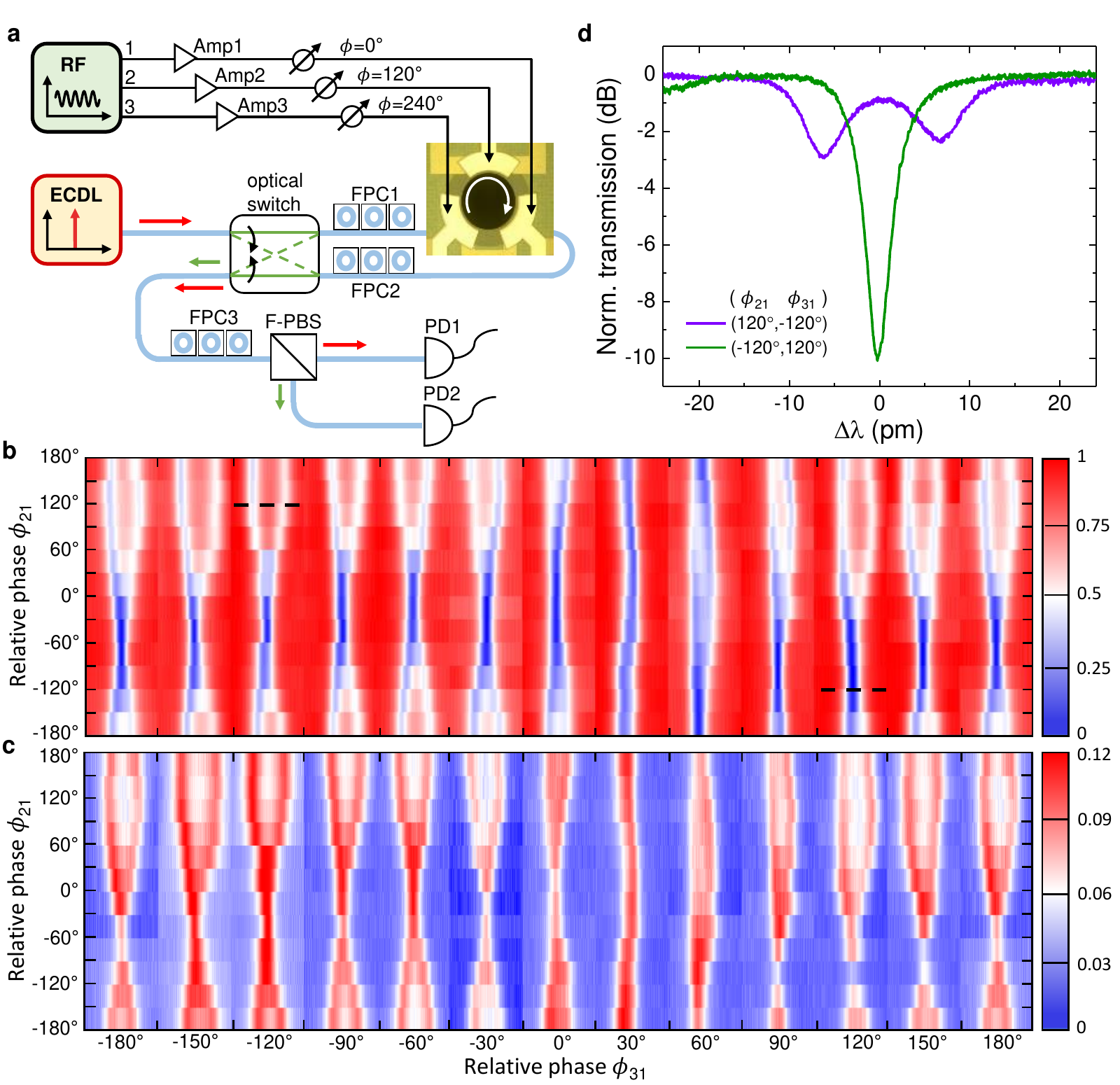}
\caption{
\textbf{Optical isolation and RF phases dependency}. 
\textbf{a} Experimental setup. 
Three RF signals are amplified and applied to the AlN actuators with phase controlled individually. 
An optical switch is used to control the direction of input TE light (red arrows). 
The output TE light (red arrows) and generated TM sideband (green arrows) are spatially separated and detected. 
The white arrow in the device denotes the clockwise rotation of the RF drive in the forward direction.
Amp: RF amplifier. 
ECDL: external-cavity diode laser.
FPC: fiber polarization controller. 
F-PBS: fiber polarization beam splitter. 
PD: photodetector. 
\textbf{b} Optical transmission and \textbf{c} converted sideband under phase sweep of signals 2 and 3 relative to signal 1, i.e. sweeping $\phi_{21}$ and $\phi_{31}$. 
Each column is an experimentally measured spectrum under the same $\phi _{31}$ with spectral span of (-16 pm, 16 pm) relative to the center wavelength $\lambda_0$ (1542.6 nm) of TE$_{00}$ mode. 
Both \textbf{b} and \textbf{c} are normalized to the input TE light power on chip. 
\textbf{d} Optical transmission spectra with $(\phi_{21}, \phi_{31})=(120^\circ, -120^\circ)$ (perfect phase matching) and $(\phi_{21}, \phi_{31})=(-120^\circ, 120^\circ)$ (largest phase mismatch). 
These two phase combinations are marked with black dashed lines in \textbf{b}. 
}
\label{Fig:3}
\end{figure*}

\noindent \textbf{Optical isolation demonstration}. 
The measurement setup is shown in Fig. \ref{Fig:3}a. 
Three RF signals are amplified and applied to each actuator, and the amplitudes and phases of each channel are controlled individually by each signal generator. 
The TE$_{00}$ mode is excited by aligning the polarization using fiber polarization controllers (FPC 1, 2). 
The light propagation direction, forward or backward, is controlled by a $2\times2$ optical switch. 
The transmitted TE light and generated TM sideband are separated by a polarization beam splitter (PBS), which is a key reason why we use two modes of different polarizations. 
The laser wavelength is continuously scanned across the resonance to probe the spectral response around the TE$_{00}$ resonance. 

The RF phases are critical for phase matching. 
Figure \ref{Fig:3}b shows the transmission spectrum of TE light by sweeping the RF phases of signals 2 and 3 relative to signal 1 ($\phi_{21}$ and $\phi_{31}$), while the output RF power (20 dBm for each actuator) and light input direction are fixed. 
Note that reversing the sign of the RF phases changes the rotation direction of the acoustic wave. 
Non-reciprocity can be seen from the disparate transmission by reversing the RF phases with respect to the origin (0$^{\circ}$, 0$^{\circ}$). 
Strong mode splitting is induced under ideal phase setting $(\phi_{21}, \phi_{31})=(120^\circ,-120^\circ)$, while the original single resonance is maintained at ($\phi_{21}, \phi_{31})=(-120^\circ,120^\circ$), as shown in Fig. \ref{Fig:3}d. 
Intriguingly, non-reciprocity only degrades slightly when the phases deviate from the ideal value, which allows large tolerance of phase settings in practical applications. 
This behaviour and RF phase dependency are also revealed by Finite-Difference Frequency-Domain (FDFD) simulations \cite{Shi:18}, showing reasonable matching with experimental data (see Supplementary Note 4). 

\begin{figure*}[t!]
\centering
\includegraphics{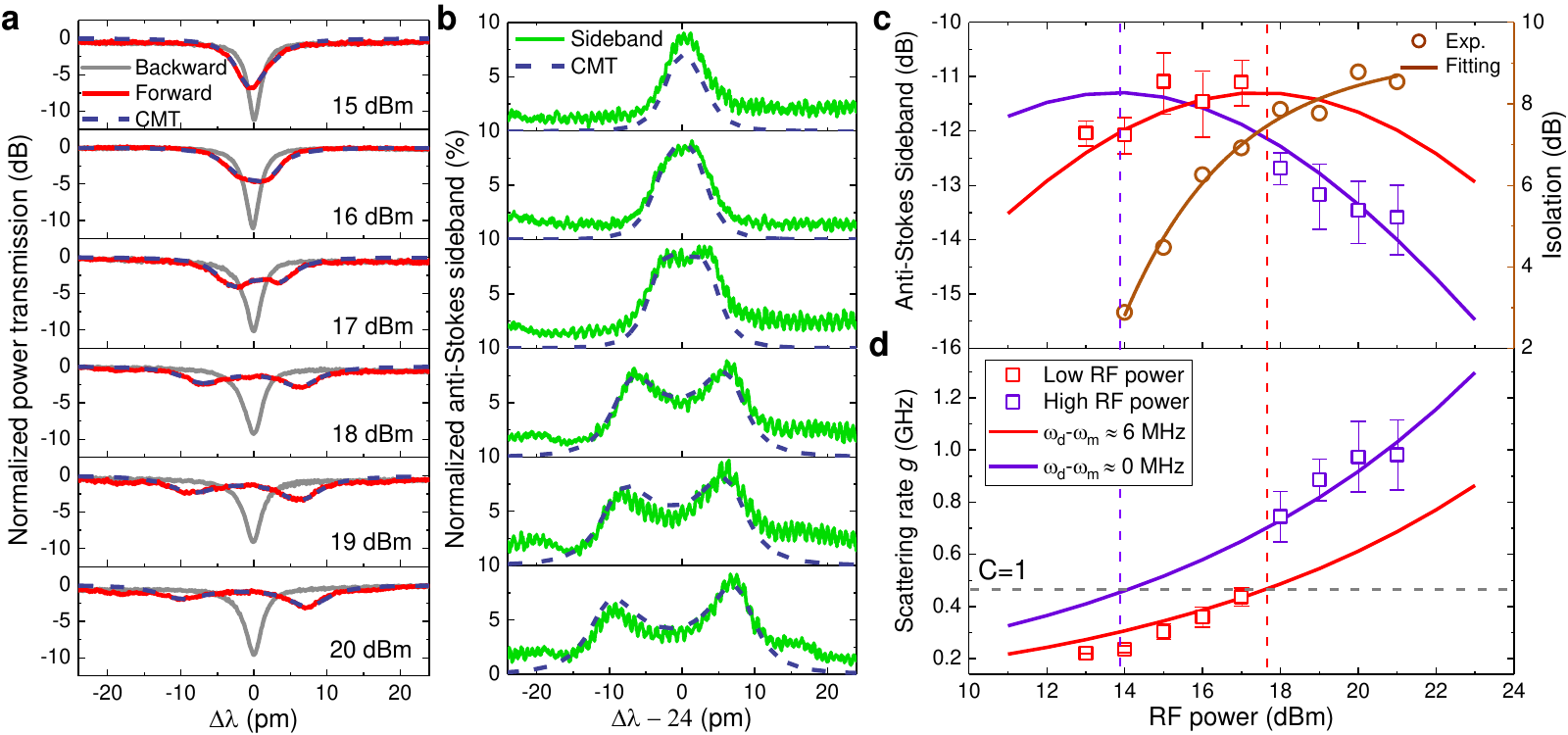}
\caption{
\textbf{RF power dependency and anti-Stokes TM sideband generation}. 
\textbf{a} Optical transmission spectra of the TE light in the forward (red solid) and backward (gray solid) directions, with increasing RF power from 15 to 20 dBm. $\Delta \lambda$ is the detuning of the input laser relative to the TE$_{00}$ mode $\Delta \lambda=\lambda_L-\lambda_{TE}$, with $\lambda_{TE}=1544.1$ nm.
\textbf{b} Generated light of the anti-Stokes TM sideband in the forward direction, normalized to the input TE light power. 
The RF power increase is same as \textbf{a}. 
The anti-Stokes sideband is blue shifted relative to the input laser by the modulation frequency 2.968 GHz ($\sim 24$ pm). 
The fitted transmission using Coupled Mode Theory (CMT, blue dashed) is also shown in \textbf{a} and \textbf{b}. 
\textbf{c} Conversion efficiency of the TM sideband at $\Delta \lambda =0$ and \textbf{d} scattering rate $g$ as a function of RF power. 
Experimental data (squares) are grouped as low RF power (red) and high RF power (purple), fitted individually with CMT (solid lines) with different microwave drive to mechanical resonance detuning $\omega _d - \omega _m$. 
Horizontal gray dashed line in \textbf{d} marks the value of $g$ (460 MHz) with cooperativity $C=1$. 
Vertical red and purple dashed lines in \textbf{c} and \textbf{d} mark the maximum conversion at each detuning. 
The error bars of each data point represent the standard deviation from 5 individual measurements. 
The dependence of isolation on RF power is shown in \textbf{c} with experimental data (brown circle) and exponential fitting (brown solid line). 
Note: the RF power in all the panels is the power applied on each individual actuator, thus the total RF power consumption is three times (4.8 dB) higher.
}
\label{Fig:4}
\end{figure*}

Light transmission of the generated anti-Stokes TM sideband is simultaneously measured as shown in Fig. \ref{Fig:3}c. 
It is normalized to the TE's input power, thus can be interpreted as conversion efficiency $\eta$. 
Figure \ref{Fig:3}b and c show similar pattern but with reversed color rendering. 
Prominent splitting and TE-TM conversion are found at $(120^\circ,-120^\circ)$, while TE-TM conversion is negligible at $(-120^\circ,120^\circ)$. 
As the measured TM sideband is resulted from mode coupling and phase matching, it can be used as feedback signal for tuning and stabilizing the RF phases.  
From Fig. \ref{Fig:3}d, the optical isolation ratio between the clockwise (forward) and counter-clockwise (backward) directions of the RF drive is calculated as 9.3 dB, which is mainly limited by the level of critical coupling (-10.1 dB) of the optical microring (the current device is slightly under-coupling). 
This can be easily improved in the future by fine tuning the bus-microring gap in the design and fabrication. 
We achieve $83\%$ transmission on resonance corresponding to 0.8 dB insertion loss in the forward direction.
We further note another device with 0.1 dB insertion loss ($98\%$ transmission) which shows higher modulation efficiency but less backward extinction due to the fact the microring is more weakly coupled with a wider gap  (see Supplementary Note 5). 

\begin{figure*}[t!]
\centering
\includegraphics{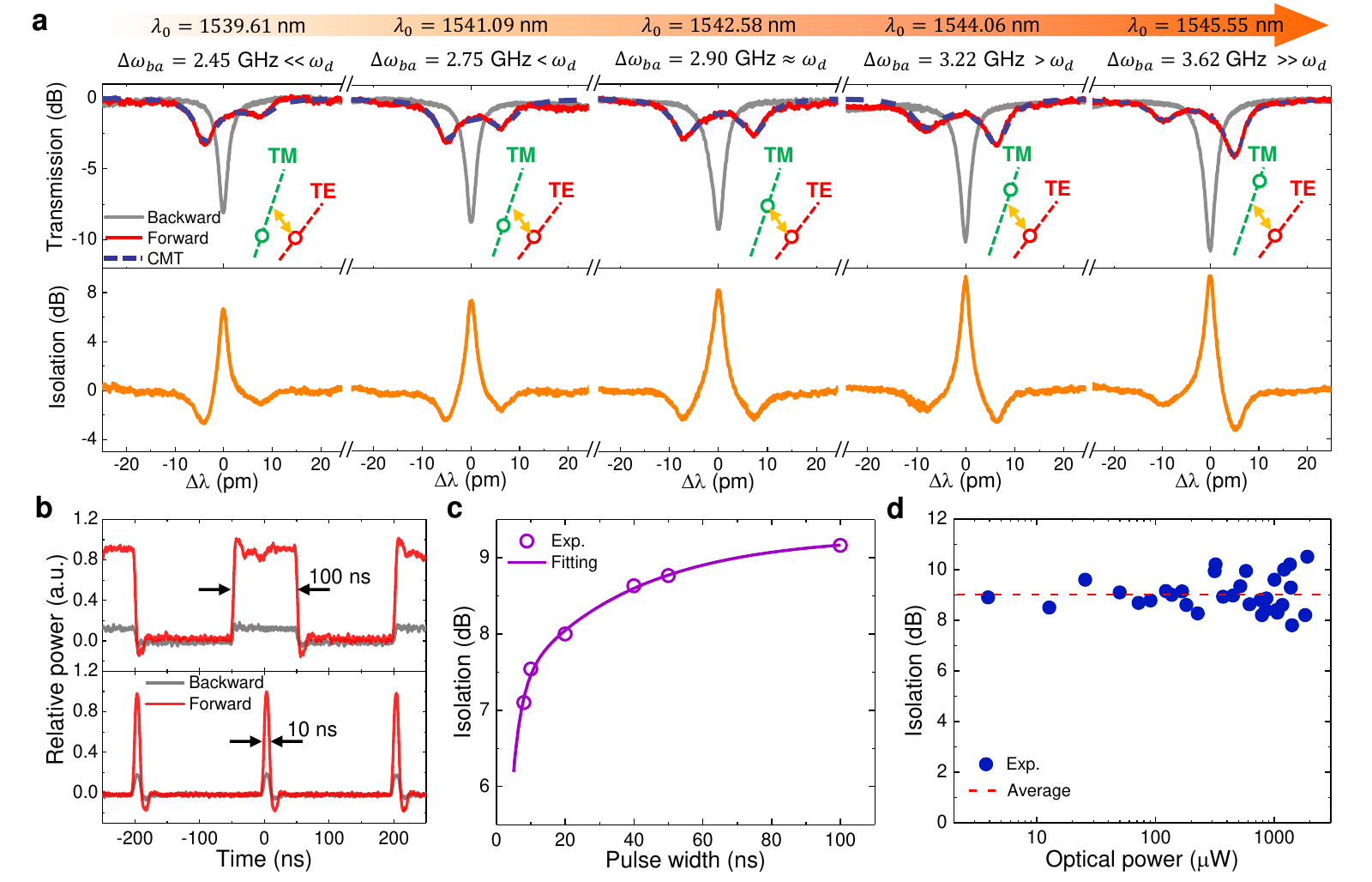}
\caption{
\textbf{Influence of optical mode spacing, time-domain response, and optical power linearity}. 
\textbf{a} Top panel: Optical transmission spectra in the forward (red) and backward (gray) directions, with increasing center wavelength (left to right). 
The mode spacing ($\Delta \omega _{ba}$) changes correspondingly due to the different FSRs of the TE$_{00}$ and TM$_{00}$ modes. 
Insets show the relative positions of the two modes in $\omega - k$ space (not to scale). 
Bottom panel: spectra of isolation ratios varying with the center wavelength. 
\textbf{b} Transmission of optical pulse trains in the forward (red) and backward (gray) directions for 100 ns (top panel) and 10 ns (bottom panel) pulse width. 
The power is normalized to the maximum power of the forward pulse in each case. 
\textbf{c} The isolation ratio decreases with narrower pulse width due to the finite photon lifetime (1.5 ns) in the optical cavity. 
Experimental data (circle) are fitted with an exponential function (solid line). 
\textbf{d} The isolation remains nearly constant around 9 dB (red dashed line) over 30 dB dynamic range of the optical input power. 
}
\label{Fig:5}
\end{figure*}

\noindent \textbf{RF power dependency}. 
The evolution of optical isolation with varying applied RF power is studied in Fig. \ref{Fig:4} with fixed RF phases of $(\phi_{21}, \phi_{31})=(120^\circ,-120^\circ)$ and drive frequency of 2.968 GHz. 
Figure \ref{Fig:4}a shows the measured forward and backward transmissions of the TE$_{00}$ mode, with the RF power (applied to each actuator) increased from 15 to 20 dBm. 
In the forward direction, initially the resonance depth decreases and the linewidth broadens with increasing RF power up to 16 dBm, resulted from the increasing intrinsic loss caused by the scattering to the TM$_{00}$ mode. 
Above 17 dBm, mode splitting appears, creating a transparency window at the original resonance frequency. 
In the backward direction, single-resonance profile remains, however with slightly increasing linewidth due to the weak mode coupling as predicted and described by the Floquet theorem \cite{Shi:18}. 
Figure \ref{Fig:4}c shows the isolation at zero detuning relative to the TE$_{00}$ mode ($\Delta \lambda =0$) which increases exponentially with the applied RF power. 

The mode splitting rate, which is two times of the interband scattering rate $g$, is extracted by fitting the resonance profile using the generalized Eq. \ref{eq3} from CMT, with $g$ and $\Delta \omega _{ba}$ being the fitting parameters. 
Figure \ref{Fig:4}d shows that $g$ gradually increases with increasing RF power, and sharply increases to a higher value at 18 dBm RF power, above which $g$ continuously increases and finally saturates at 20 dBm. 
This behaviour is caused by the blue shift of the SiO$_2$ HBARs due to RF thermal heating, as SiO$_2$ has a large positive temperature coefficient of elasticity \cite{rais:2013} of 188 ppm K$^{-1}$. 
The RF drive frequency $\omega _d$ is initially blue-detuned from the HBAR frequency $\omega _m$ at room temperature, i.e. $\omega _d-\omega _m>0$. 
As the RF power is increased and the acoustic velocity in SiO$_2$ increases, $\omega _m$ approaches $\omega _d$ (i.e. $\omega _d-\omega _m \rightarrow 0$) and more phonons are pumped into the mechanical cavity. 
This in turn increases the temperature which further blue-shifts the HBAR. 
Thus, the thermal nonlinearity leads to an increase in $g$ at an RF power of approximately 18 dBm. 
This transition is studied in Supplementary Note 7 with fine sweep of RF power in 0.1 dBm step.  

With RF power above 20 dBm, the interband scattering rate $g$ saturates, which suggests the efficiency of pumping phonons into the cavity starts dropping. 
This is likely because $\omega _d$ becomes red-detuned to $\omega _m$, i.e. $\omega _d-\omega _m<0$. 
Therefore, for RF power between 18 to 20 dBm, nearly zero detuning can be derived. 
In this regime, the single-phonon optomechanical coupling strength $g_0$ is calculated as 208 Hz, by fitting high RF power data (purple line) with $g=g_0 \sqrt{n_c}$, where $n_c$ is calculated by extracting the electro-mechanical coupling efficiency $k_\text{t,eff}^2 =0.2\%$ from S$_{11}$. 
The low RF power region is fitted with $\omega _d-\omega _m\approx6$ MHz blue-detuning (red line). 
Another consequence of the RF thermal effect is the drift of mode spacing $\Delta \omega _{ba}$. 
Figure \ref{Fig:4}a shows that the mode splitting evolves from symmetric (18 dBm) to asymmetric (20 dBm) with increasing RF power, and $\Delta \omega _{ba}$ is increased from 3 GHz to 3.3 GHz. 

The spectrum of the TM anti-Stokes sideband is shown in Fig. \ref{Fig:4}b. 
The conversion efficiency $\eta$ when the input laser is zero-detuned to the TE$_{00}$ mode ($\Delta \lambda =0$) is plotted in Fig. \ref{Fig:4}c. 
Similarly, each data set with low and high RF power are fitted individually using Eq. \ref{eq4} with the same detuning as used in Fig. \ref{Fig:4}d. 
It can be derived from Eq. \ref{eq4} that the maximum value of $\eta$ is reached at $C=1$. 
This can also be seen from Fig. \ref{Fig:4}b where the conversion starts to drop at center due to the mode splitting when $C>1$. 
Maximum of $8 \%$ (-11 dB) of the TE$_{00}$ mode power is converted to the TM$_{00}$ sideband, which is mainly limited by the external coupling efficiency of the TE$_{00}$ ($\kappa _{a,\text{ex}} / \kappa _a =0.34$) and TM$_{00}$ ($\kappa _{b,\text{ex}} / \kappa _b =0.24$) modes (see Eq. \ref{eq4}). 
At zero RF drive detuning, $C=1$ is achieved with 14 dBm RF power applied on each actuator (18.8 dBm in total), and the system lies in the strong coupling regime at 20 dBm. 
On the other hand, the detuning not only lowers $g$ but increases the required RF power to achieve $C=1$. 
Finally, it is worth noting that the same device can work as a non-reciprocal frequency shifter \cite{hu:2020} with $100 \%$ conversion achievable for device that is strongly over-coupled (i.e. $\kappa _\text{ex} \approx \kappa$). 
It could serve as a key building block in photonic quantum computing \cite{Kobayashi:16, Lukens:17, Joshi:20} and microwave photonics \cite{Marpaung:19}. 

\begin{table*}[t]
    \centering
    \caption{Comparison of monolithically integrated, magnetic-free optical isolator devices. }
    \begin{tabular}{p{1.2cm}p{2.9cm} p{1.5cm}p{1.8cm}p{1.5cm}p{2cm}p{1.7cm}p{1.5cm}  }
 \hline
  Year & Scheme & Structure & Material & Isolation & Insertion loss & Bandwidth & Power \\
 \hline
2020\cite{YangKY:20} & Nonlinear optics & Ring & Si & 20 dB & 1.3 dB& 20 GHz& No drive \\
2014\cite{Tzuang:14} & Synthetic magnetic & MZI &  Doped Si & 2.4 dB & & 20 nm& 34 dBm \\
2021\cite{Kim:21} & Synthetic magnetic & Ring &  AlN & 3 dB & 9 dB& 4 GHz& 16 dBm \\
2021\cite{Dostart:21} & Synthetic magnetic & Ring &  Doped Si & 13 dB & 18 dB& 2 GHz& -3 dBm \\
2012\cite{Lira:12} & Spatio-temporal & MZI &  Doped Si & 3 dB & 70 dB& 200 GHz& 25 dBm \\
2018\cite{Kittlaus:18} & Spatio-temporal$^a$ & MZI &   Si & 39 dB & NA & 125 GHz& 90 mW$^b$\\
2018\cite{Sohn:18} & Spatio-temporal$^a$ & Ring &  AlN & 15 dB & NA& 1 GHz& 18 dBm \\
2021\cite{Kittlaus:21} & Spatio-temporal$^a$ & MZI &  Si+AlN & 16 dB & NA& 100 GHz& 21 dBm \\
This work & Spatio-temporal & Ring &  Si$_3$N$_4$+AlN & 10 dB & 0.1-1 dB& 0.7 GHz& 20 dBm \\
 \hline
\end{tabular}
\label{table1}
\footnotetext{Non-reciprocal sideband modulation} \footnotetext{Optical drive}
\end{table*}

\noindent \textbf{Detuning of optical mode spacing}. 
Since there is 380 MHz difference in FSR between the TE$_{00}$ and TM$_{00}$ modes (see Supplementary Note 2), their frequency spacing $\Delta \omega_{ba}$ varies from pair to pair for different center wavelengths.
Here, the dependence of isolation performance on $\Delta \omega_{ba}$ is studied in Fig. \ref{Fig:5}a under 20 dBm RF power. 
At $\lambda _0=1542.58$ nm, $ \Delta \omega_{ba}$ is nearly equal to the driving frequency $\omega _d$, and the mode splitting is symmetric. Because the optical resonance linewidths are around gigahertz level, TE and TM modes can still be coupled even with a frequency mismatch between $\Delta \omega _{ba}$ and $\omega _d$ on the order of 0.5 GHz. 
The mismatch leads to asymmetric mode splitting. 
Nevertheless, there is no prominent degradation of isolation within the measured range. 
The decrease of maximum isolation for shorter wavelength is caused by the reduction of extinction in the backward direction. 
This is because that shorter wavelength has smaller mode size and thus a weaker bus-microring external coupling rate $\kappa_\text{ex}$ that leads to under-coupling. 
Maximum of 9.5 dB isolation is achieved at 1545.55 nm, which is even larger than the matched symmetric case due to its better critical coupling. 
Therefore, the isolator can work simultaneously for multiple center wavelengths which is important for optical communication using wavelength multiplexing \cite{Marin-Palomo:17}. 
The wavelength range can be extended to cover the optical C-band by using pulley coupling scheme \cite{Moille:19a} to maintain critical coupling or slight over-coupling, and engineering the waveguide geometry to reach precise FSR match. 
On the other hand, thermal tuning to continuously shift the operating resonance can further increase the bandwidth. 

\noindent \textbf{Isolation of optical pulses}. 
We further evaluated the optical isolator to demonstrate a unidirectional transmission of optical pulse train to mimic (0, 1) data stream, as shown in Fig. \ref{Fig:5}b. 
The 100-ns pulse shows the quasi-static response, where the backward reflection is only $11\%$ of the forward transmission. 
The bump at the pulse edge is caused by the limited bandwidth (125 MHz) of the photodetector. 
The dynamic response is tested with 10-ns pulses, illustrating a vast contrast in the two directions. 
The isolation bandwidth can be inferred from the decreasing isolation with decreasing pulse duration, as revealed in Fig. \ref{Fig:5}c. 
A pulse minimum of 8 ns is measured, limited by the photodetector’s bandwidth. 
Over 8 dB isolation is maintained for pulses longer than 20 ns. 
However, the isolation drops exponentially when the pulse duration is shorter than 20 ns, which is ultimately limited by the photon lifetime of $\sim1.5$ ns, corresponding to 680 MHz linewidth of the optical resonance. 

\noindent \textbf{Optical power linearity}. 
Dynamic reciprocity has been a well-known limitation for most optical isolators relying on optical nonlinearity, where the isolation degrades dramatically when light transmits simultaneously in both directions with optical power exceeds the threshold \cite{Shi:15, YangKY:20}. 
In our work, with only an electrical drive, the optical linearity is preserved as long as the intra-cavity photon number is smaller than the phonon number. 
Since phonon frequency is five orders of magnitude smaller than photon frequency, theoretically it suggests maximum of 6 kW optical power for the 20-dBm RF power in the experiment. 
However, the optical linearity is finally limited to several Watts due to the Kerr nonlinearity of Si$_3$N$_4$. 
The quasi-square waveguide used here has a normal group velocity dispersion (GVD), thus Kerr parametric oscillation is suppressed. 
The linearity is experimentally verified in microwatt to milliwatt range as shown in Fig. \ref{Fig:5}d. 
The optical isolation remains nearly constant within the measurement range. 
The large variation at high optical power is mainly due to the optical thermal nonlinearity as we sweep the laser across the optical resonance. 
Theoretically there is no limit on the lower bound of optical power which can work for photonic quantum computing \cite{Kobayashi:16, Lukens:17, Joshi:20}. 

\section{Discussion}
This work has demonstrated an integrated optical isolator achieved by spatio-temporal modulation of a Si$_3$N$_4$ microring resonator via three AlN pieozoelectric actuators. 
The device has been fully characterized in terms of RF phases, RF powers, optical spectra, and optical powers, showing agreement with theoretical models and numerical simulations. 
Table \ref{table1} summarizes recently demonstrated integrated, magnetic-free isolators (to the best of our knowledge), in comparison with our work. 
A comprehensive comparison with other experimental realizations can be found in Supplementary Note 8. 

Although promising progresses have been made using nonlinear optics \cite{YangKY:20} and synthetic magnetic field \cite{Kim:21, Dostart:21}, most spatio-temporal modulation demonstrations \cite{Kittlaus:18, Sohn:18, Kittlaus:21} are still in the non-reciprocal sideband modulation regime. 
In this work, the HBAR AOM helps to boost the optical cooperativity $C$ beyond 1 and enter strong coupling regime, thanks to its power handling capability and the tight acoustic confinement. 
We obtained an isolation of 10 dB under an RF power of 20 dBm. 
Notably, a record-low insertion loss of only 0.1 dB is achieved due to the intrinsic low loss of Si$_3$N$_4$ waveguides. 
In contrast to nonlinear optics, the electrical drive largely preserves the optical linearity by separating RF driving and optical sensing in two different domains.

In the future, the isolation can be significantly improved by optimizing the bus-microring coupling gap to reach critical coupling or slight over-coupling \cite{Pfeiffer:17b}. 
Furthermore, it has been theoretically predicted that the RF power consumption can be reduced to a few dBm by reducing the microring radius to 20 $\upmu$m, which will increase both the electro-mechanical and optomechanical coupling efficiencies \cite{Blesin:21}. With these improvements, electrically driven, magnetic-free optical isolator could be reliably incorporated as a key building block into current integrated opto-electronic systems. 

\noindent\textbf{Methods}

\medskip

\begin{footnotesize}

\noindent \textbf{Device fabrication}:
The Si$_3$N$_4$ PIC is fabricated using the photonic Damascene process \cite{Pfeiffer:18b, Liu:18a, Liu:20b}. The monolithic integration of piezoelectric actuators on top of Si$_3$N$_4$ microresonators is illustrated in  Ref. \cite{Tian:20, Liu:20a}. 
Films of 100 nm Mo and 1 $\upmu$m AlN are sputtered on the wafer through foundry services (Plasma-Therm). 
The actuators are patterned by thick photoresist SPR220-4.5 and dry etched using Cl$_2$ and BCl$_3$ in Panasonic E620 Etcher. 
The bottom Mo electrodes are patterned by photoresist AZ1518 and dry etched using Cl$_2$ in the same etcher. 
Finally, the top 100 nm Al is deposited by an electron-beam evaporator and patterned using a standard lift-off process. 
The SiO$_2$ release process is shown in Fig. \ref{Fig:1}d. 
The center release hole is patterned using photolithography, and SiO$_2$ is dry-etched to expose the Si substrate. 
The Si is then isotropically etched using SF$_6$ Bosch process to undercut and suspend the SiO$_2$ membrane.  

\noindent \textbf{Design of Si$_3$N$_4$ waveguides for phase matching}: 
From the microring's resonant condition $k=2\pi /\lambda =m/R$, the momentum is related to the azimuthal order $m$ of the mode and the microring's radius $R$.
As the three actuators cover the entire microring, the generated rotating acoustic wave has an effective wavelength of $2\pi R$, and thus an azimuthal order of 1 ($m_c=1$). 
Thus the phase matching condition requires the azimuthal order difference $\Delta m_{ab}=m_a-m_b=m_c=1$ between the TE$_{00}$ and TM$_{00}$ modes.
As the Si$_3$N$_4$ waveguide is surrounded by isotropic SiO$_2$ cladding, a quasi-square waveguide cross-section ($810 \times 820$ nm$^2$), as shown in Fig. \ref{Fig:2}b inset, is designed to have slightly different effective refractive indices for the two optical modes (see Supplementary Note 1). 
Here, phase matching requires two optical modes to co-propagate, with counter-propagating acoustic wave. 
In the final devices used in the experiment, the azimuthal order difference $\Delta m_{ab}$ is measured and calibrated (see Supplementary Note 2) to be around 4.  
Due to the discrete nature of the spatial modulation, there are higher-azimuthal-order Fourier components simultaneously excited that can fulfil phase matching at the expense of lower efficiency \cite{Sounas:14}.

\noindent \textbf{Measurement setup}: 
The electromechanical S$_{11}$ is measured by detecting the reflected microwave signal using a vector network analyzer (VNA, Agilent E8364B), where the electrical signal is applied to the device through an RF GSG probe (Cascade ACP40-GSG-150).  
To measure the optomechanical S$_{21}$, continuous-wave (CW) light from a diode laser (Velocity Tunable Laser 6328) is edge-coupled into the chip using a lensed fiber and an inverse taper with around 50 $\upmu$W power on chip. 
An RF signal of -5 dBm power is applied from the port 1 of the VNA to drive the AlN actuator and the light intensity modulation is detected by a 12-GHz photodiode (New Focus 1544), which is sent back to port 2 of the network analyzer.

In the main optical isolation measurement, three RF signal generators (Agilent E8257D) output RF drive for each actuator and are amplified by three RF amplifiers (ZHL-5W-63-S+) before going into the device. The phases between them are controlled by their built-in phase controller after they are synchronized by their internal 10 MHz clock. The same diode laser outputs CW light whose wavelength is continuously swept around the optical resonance. The input polarization is controlled by fiber polarization controllers (FPC561) to align with the TE polarization on chip. The light input direction is selected by a MEMS (micro-electromechanical system) $2\times 2$ optical switch (OSW22-1310E). The output TE and TM polarized light is separated by a fiber polarization beam splitter (PBC1550SM-FC) and measured by two photodetectors (New Focus 1811), whose signals are recorded by an oscilloscope (MSO8104A). Note the optical axis of the beam splitter is fixed, thus we have to use a polarization controller to align with it. During the measurement, TE polarization is selected by rotating the polarization controller until only TE mode resonance is observed. This process is repeated for both directions. It is worth noting that, although the polarization will change as we switch the MEMS optical switch, the polarization right before the PBS will be the same, since the optical switch itself is a reciprocal device. 

For testing the isolation of optical pulses, the input light before the optical switch is modulated by an electro-optical intensity modulator (Lucent 2623CSA) and the input pulse is generated by a function generator (Agilent 33250A). In testing of the optical linearity, Erbium Doped Fiber amplifier (EDFA-I-B) is used before the optical switch for optical powers larger than 100 $\upmu$W. 

\noindent \textbf{Derivation of coupled mode equations}: 
The three-wave mixing (two optical modes $a$ and $b$ and one mechanical mode $c$) process can be described by the quantum interaction Hamiltonian as:
\begin{equation}
    H_I=\hbar g_0(ab^{\dag } c+a^{\dag }bc^{\dag })
\end{equation}
assuming phase matching is fulfilled and $a$ has smaller frequency than $b$. Thus, only two processes can be supported: 1. $ab^{\dag } c$, annihilation of photon $a$ and phonon $c$ and generation of one photon $b$; 2. $a^{\dag }bc^{\dag }$, annihilation of one photon $b$ and generation of photon $a$ and phonon $c$. Following similar approach as Ref \cite{Fan:18}, the equation of motion can be obtained by assuming resolved sideband and rotating-wave approximation:
\begin{eqnarray}
\label{eq6}
\dot{a} & = & -(i\Delta _a +\frac{\kappa _{a}}{2})a-ig_0 bc^{\dag} +\sqrt{\kappa _{a,\text{ex}}}a_\text{in}\\
\label{eq7}
\dot{b} & = & -(i\Delta _b +\frac{\kappa _{b}}{2})b-ig_0 ac\\
\label{eq8}
\dot{c} & = & -(i\omega _m +\frac{\gamma _{c}}{2})c-ig_0 a^{\dag}b +\sqrt{\gamma _{c,\text{ex}}}c_\text{in}e^{-i\omega _dt}\\
a_{\text{out}} & = & a_{\text{in}}-\sqrt{\kappa _{a,\text{ex}}}a\\
b_{\text{out}} & = & -\sqrt{\kappa _{b,\text{ex}}}b
\end{eqnarray}
where $\gamma _{c}$ and $\gamma _{c,\text{ex}}$ are total loss rate and external coupling rate of the mechanical mode, and $c_\text{in} = \sqrt{P_\text{in}/\hbar \omega _d}$ is input microwave amplitude. Assuming optomechanical back-action (term $g_0 a^{\dag}b$ in Eq. \ref{eq8}) is much smaller than the microwave drive, the intra-cavity amplitude of $c$ at static state is:
\begin{eqnarray}
\label{eq11}
c & = & \sqrt{n_c}e^{-i\omega _dt}\\
n_c & = & \frac{\gamma _{c,\text{ex}}}{(\omega _d - \omega _m)^2 +\gamma _{c}^2/4}\frac{P_\text{in}}{\hbar \omega _d}
\end{eqnarray}
By inserting Eq. \ref{eq11} into Eqs. \ref{eq6}, \ref{eq7}, we will get Eqs. \ref{eq1}, \ref{eq2} in the main text. Due to the modulation, $b$ is frequency shifted by $\omega _d$ in the rotating frame of $\omega _L$. The slow amplitude and fast oscillation of $b$ can be separated by substitute $b$ with $\Tilde{b}e^{-i\omega _d t}$. At steady state, Eqs. \ref{eq1}, \ref{eq2} can be solved by letting time derivatives to be zero, and after some linear algebra we will arrive at the general expressions for transmission $T$ and sideband conversion efficiency $\eta$:
\begin{equation}
    T  =  \left| \frac{a_\text{out}}{a_\text{in}} \right| ^2  =  \left| 1- \frac{\kappa _{a,\text{ex}}}{i\Delta _a+\frac{\kappa _a}{2}+\frac{g^2}{i(\Delta_b-\omega_d)+\frac{\kappa_b}{2}}} \right| ^2
    \label{eq13}
\end{equation}
\begin{equation}
    \eta  =  \left| \frac{b_\text{out}}{a_\text{in}} \right| ^2  =  \frac{\kappa _{a,\text{ex}}}{\kappa _a}\frac{\kappa _{b,\text{ex}}}{\kappa _b}\frac{4C}{|C+(1+\frac{2i\Delta_a}{\kappa_a})(1+\frac{2i(\Delta_b-\omega_d)}{\kappa_b})|^2}
    \label{eq14}
\end{equation}
where $\Delta_b-\omega_d=\Delta_{ba}+\Delta_a-\omega_d$. It can be seen when $\Delta_a=0$, $\Delta_{ba}=\omega_d$, Eqs. \ref{eq13}, \ref{eq14} reduce to Eqs. \ref{eq3}, \ref{eq4}. Eqs. \ref{eq13}, \ref{eq14} are used for fittings in Fig. \ref{Fig:4} and \ref{Fig:5}. The mode splitting can be found from Eq. \ref{eq13} if assuming $\Delta_a \gg \kappa _a, \kappa _b$ and $\Delta_{ba}=\omega_d$:
\begin{equation}
    T  \approx \left| 1- \frac{(i\Delta_a+\frac{\kappa_b}{2})\kappa _{a,\text{ex}}}{(g+\Delta_a)(g-\Delta_a)+i(\frac{\kappa_a+\kappa_b}{2})\Delta_a} \right| ^2
    \label{eq15}
\end{equation}
where the transmission minima is reached when $\Delta_a=\pm g$. Thus the mode splitting is $2g$. 

\noindent \textbf{Acknowledgments}:
This work was supported by U.S. National Science Foundation’s RAISE TAQS program under grant PHY 18-39164, by NSF QISE-Net under grant DMR 17-47426, 
by the Air Force Office of Scientific Research under Award No. FA8655-20-1-7009,
by funding from the EU H2020 research and innovation programme under grant agreement No. 732894 (HOT), 
and by Swiss National Science Foundation under grant agreement No. 176563 (BRIDGE).
Samples were fabricated in the EPFL center of MicroNanoTechnology (CMi), and Birck Nanotechnology Center at Purdue University.
AlN deposition was performed at Plasma Therm LLC.

\noindent \textbf{Author contribution}: 
H.T. and J.L. designed the devices. 
J.L., H.T. and R.N.W. developed the process and fabricated the samples, with the assistance from J.H.. 
H.T. performed the experiment and simulations, and analyzed the data, with the assistance from A.S. and T.B.. H.T. and J.L. wrote the manuscript, with the input from others. 
S.A.B and T.J.K supervised the collaboration.

\noindent \textbf{Data Availability Statement}: The code and data used to produce the plots within this work will be released on the repository \texttt{Zenodo} upon publication of this preprint.

\end{footnotesize}
\bibliographystyle{apsrev4-1}
\bibliography{bibliography}
 \end{document}


\title{Supplementary Information: Magnetic-Free Silicon Nitride Integrated Optical Isolator}  

\author{\noindent Hao Tian$^{1,*}$, Junqiu Liu$^{2,*}$, Anat Siddharth$^{2}$, Rui Ning Wang$^{2}$, Terence Blésin$^{2}$, Jijun He$^{2}$, \\ Tobias J. Kippenberg$^{2,\dagger}$, Sunil A. Bhave$^{1,\dagger}$}
\maketitle

\begin{affiliations}
\item OxideMEMS Lab, Purdue University, 47907 West Lafayette, IN, USA
\item Institute of Physics, Swiss Federal Institute of Technology Lausanne (EPFL), 1015 Lausanne, Switzerland

\normalsize{$*$ These authors contributed equally to this work.}

\normalsize{$^\dagger$ E-mail: tobias.kippenberg@epfl.ch, bhave@purdue.edu}
\end{affiliations}

\clearpage
\spacing{1}

\section*{Supplementary Note 1: Design of quasi-square optical waveguide}

\begin{figure*}[htbp]
\centering
\includegraphics[width=\textwidth]{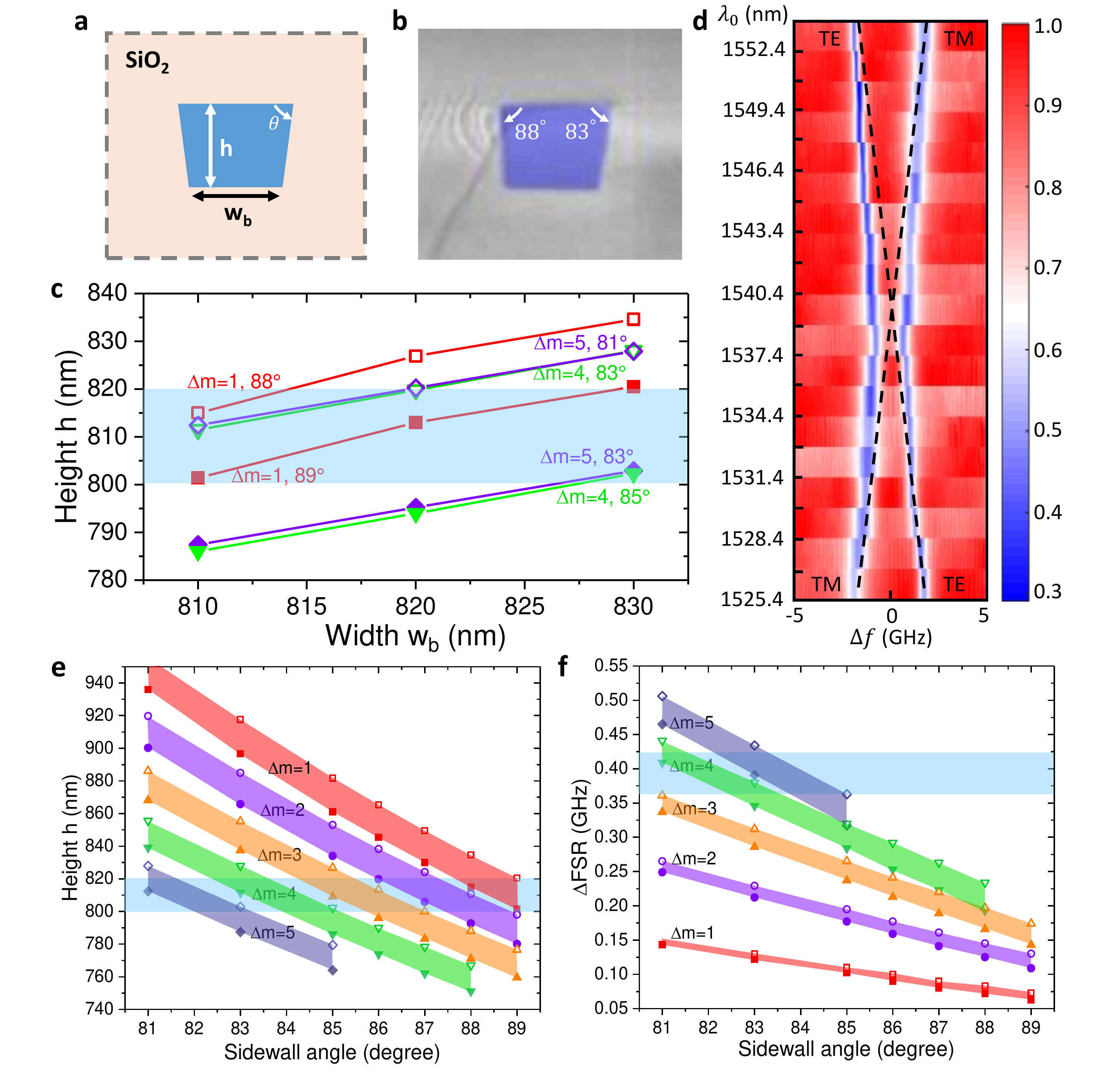}
\caption{\textbf{Optical waveguide design and calibration of azimuthal order difference $\Delta m$}. \textbf{a} Schematic of optical waveguide cross-section with Si$_3$N$_4$ (blue) embedded in isotropic SiO$_2$ cladding. The definition of height h, bottom width w$_\text{b}$, and sidewall angle $\theta$ are as labeled. \textbf{b} False colored SEM showing the cross-section of the fabricated optical waveguide. The sidewall angles are as labeled. \textbf{c} The simulated required h and w$_\text{b}$ combination that will produce close TE$_{00}$ and TM$_{00}$ modes around 1550 nm with $\Delta m=1$ (red), $\Delta m=4$ (green), $\Delta m=5$ (purple). Different sidewall angles are studied and labeled. The blue shaded area indicates the variation of waveguide height in the fabrication. \textbf{d} Measured TE$_{00}$ and TM$_{00}$ mode anti-crossing around 1540 nm. The wavelength $\lambda_0$ is the center wavelength between E$_{00}$ and TM$_{00}$. \textbf{e} Simulated height h that generates close TE$_{00}$ and TM$_{00}$ modes and \textbf{f} corresponding FSR difference $\Delta \text{FSR}$ for different sidewall angles and $\Delta m$. The variation of width w$_\text{b}$ is included as lower (810 nm) and higher (830 nm) bounds of each $\Delta m$ plot. The blue shaded areas are experimental variations. 
}
\label{SM:1}
\end{figure*}

As mentioned in the main text, satisfying phase matching condition is key for efficient modulation and mode coupling. This requires the two optical modes have frequency difference equals the mechanical frequency (below 5 GHz), and azimuthal order difference equals the effective azimuthal order of the modulation wave. It is easy to satisfy frequency requirement since at the mode anti-crossing of any two mode families we will have close spacing. It is more stringent for momentum. Since the three actuators cover the whole ring, the effective wavelength of the modulation wave is $2\pi R$ which indicates the azimuthal order is 1. This requires the two optical modes to have very close effective refractive index. In the case Si$_3$N$_4$ is embedded in isotropic SiO$_2$ cladding, an intuitive way is to design a quasi-square waveguide where the difference of effective refractive index between the two fundamental TE$_{00}$ and TM$_{00}$ modes is dictated by the aspect ratio of the waveguide's height h and width w$_\text{b}$, as shown in Fig. \ref{SM:1}a. The sidewall angle $\theta$ is also included in the design based on previous experimental experience \cite{Pfeiffer:18b, Liu:18a, Liu:20b}. As h depends on film deposition, w$_\text{b}$ on photo-lithography, $\theta$ on Reactive Ion Etching (RIE), they all show variations due to fabrication, which should be taken into account in the waveguide design.

For specific waveguide geometry (h, w$_\text{b}$, $\theta$), the resonant frequency and azimuthal order of the two modes are simulated using Finite Element Method (COMSOl), where the frequency and azimuthal order difference $\Delta m = m_\text{TE} - m_\text{TM}$ can be calculated. For fixed choice of $\theta$ and $\Delta m$, there is a combination of h and w$_\text{b}$ that will generate mode anti-crossing around wavelength of 1550 nm (frequency of 193.5 THz), as can be seen in Fig. \ref{SM:1}c. The initial design targets at $\Delta m=1$ with sidewall angle $\theta=89^\circ$ (red line with red filled square). As the waveguide height varies between 800--820 nm across the wafer, the waveguide width is designed to vary from 810 to 830 nm to take into account the fabrication variations. The width is designed to be slightly larger than height such that TE$_{00}$ has higher effective refractive index than TM$_{00}$, and thus slightly higher azimuthal order. To maintain $\Delta m$, width and height needs to increase (or decrease) together to keep the aspect ratio. As the sidewall angle decreases to $88^\circ$, the required height increases at the same width (red line with red unfilled square). Nevertheless, the curve still crosses the height variation (blue shaded area) at low width. In this case, the initial design is supposed to cover the variation of height, width and sidewall angle. 

\section*{Supplementary Note 2: Calibration of azimuthal order difference $\Delta m$}
In this section, the azimuthal order difference $\Delta m$ of the fabricated device is calibrated, showing slight deviation from the initial design in last section. At first, the cross-section of the waveguide can be seen from the SEM in Fig. \ref{SM:1}b. The sidewall angles deviate from $89^\circ$ with smallest angle being $83^\circ$, which will be corrected in the following analysis. For the device that is measured in the main text, the spectra of TE and TM mode pairs are measured over a broad wavelength range, as shown in Fig. \ref{SM:1}d. We found there is mode anti-crossing around 1540 nm. At long wavelength, TE mode falls behind TM mode in frequency $f_\text{TE}<f_\text{TM}$, while at short wavelength TE shows higher frequency $f_\text{TE}>f_\text{TM}$. This suggests TE mode has higher FSR than TM mode, such that $\Delta \text{FSR} = \text{FSR}_\text{TE} - \text{FSR}_\text{TM}\approx 380$ MHz. This FSR difference is used to find approximately the $\Delta m$ of the measured device by comparing with numerical simulations (COMSOL), as illustrated in Fig. \ref{SM:1}f. 

Figure \ref{SM:1}e simulates the required h as a function of sidewall angle for different $\Delta m$. The variation of width is also included in each $\Delta m$ plot with the lower bound being $w_\text{b}=810$ nm and upper bound being $w_\text{b}=830$ nm. From the figure we can find as the sidewall angle decreases, the required height increases, and it is higher for smaller $\Delta m$. For angle around $83^\circ$, the height achieved in the experiment overlaps with $\Delta m=4\sim 5$. In Fig. \ref{SM:1}f, the change of $\Delta \text{FSR}$ with sidewall angle is similarly plotted. It can be seen $\Delta \text{FSR}$ is smaller for small $\Delta m$, which means the two optical modes has closer index. $\Delta \text{FSR}$ also increases with decreasing sidewall angle. Based on the measured $\Delta \text{FSR}$, $\Delta m=4$ shows better overlap with both height and $\Delta m$. In the future, $\Delta m$ can be made as close as 1 with better estimation of the variations of waveguide's height, width, and sidewall angles.

\section*{Supplementary Note 3: Electromechanical model of HBAR mode}
\begin{figure*}[h]
\centering
\includegraphics[width=\textwidth]{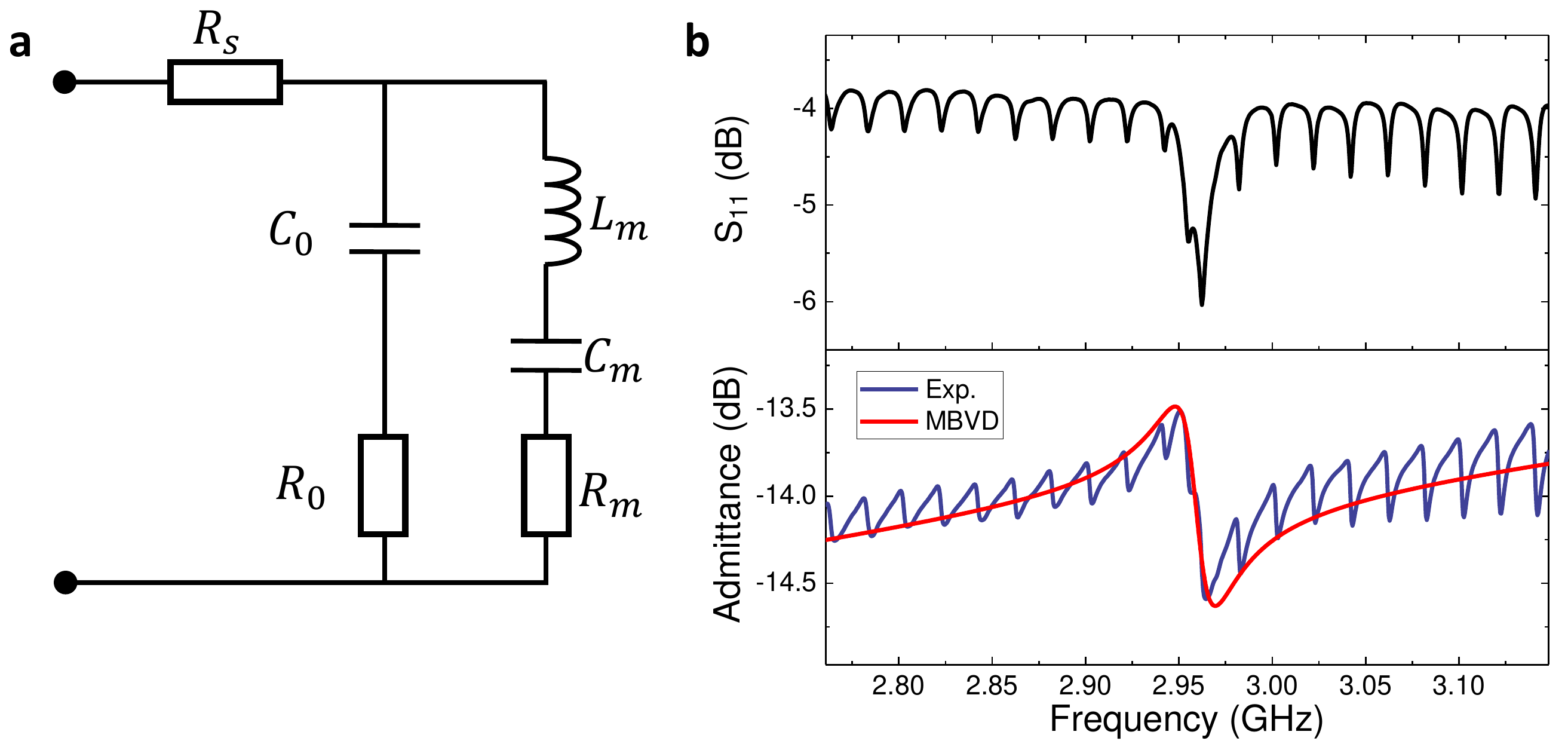}
\caption{\textbf{Electromechanical MBVD model fitting with the HBAR mode}. \textbf{a} Schematic of equivalent electrical circuit model representing one HBAR mode. \textbf{b} Upper panel: electromechanical S$_{11}$ around the HBAR mode used in the experiments in the main text. Lower panel: Calculated admittance (blue) and MBVD fitting (red). 
}
\label{SM:2}
\end{figure*}

The HBAR that is used in the experiments in the main text is analyzed using the well-known Modified Butterworth-Van Dyke model (MBVD) \cite{larson:2000}, which describes the HBAR resonance by an equivalent $RLC$ circuit as shown in Fig. \ref{SM:2}a. $L_m$ and $C_m$ are motional inductance and capacitance which present kinetic and potential energy of the mechanical resonance, while the motional resistance $R_m$ accounts for the intrinsic mechanical loss \cite{Blesin:21}. The mechanical $R_m L_m C_m$ branch is parallel to a capacitance $C_0$ which is the physical capacitance formed by the top and bottom electrodes of the piezoelectric actuator. $R_0$ is the loss of the capacitor and $R_s$ is the series resistance. The admittance looking into the whole circuit can be calculated as \cite{larson:2000}:
\begin{equation}
    Y(\omega)=j\omega C_0 \frac{1-(\frac{\omega}{\omega_p})^2+j(\frac{\omega}{\omega_p})\frac{1}{Q_{p0}}}{1-(\frac{\omega}{\omega_s})^2+j(\frac{\omega}{\omega_s})\frac{1}{Q_{s0}}}
    \label{eq1}
\end{equation}
where $\omega_s=1/\sqrt{L_m C_m}$ is the series resonance of the $R_m L_m C_m$ branch, which is also the mechanical resonance. $\omega_p$ is the so-called parallel resonance, which is related with $\omega_s$ by :
\begin{equation}
    \left(\frac{\omega_p}{\omega_s}\right)^2=1+\frac{C_m}{C_0}
\end{equation}
$Q_{s0}$ and $Q_{p0}$ are the $Q$ of each resonance:
\begin{eqnarray}
    \frac{1}{Q_{s0}} &=& \omega_s(R_m+R_s)C_m  \\
    \frac{1}{Q_{p0}} &=& \omega_p(R_0+R_m)C_m
\end{eqnarray}

The S$_{11}$ of the HBAR mode around 3 GHz is shown in Fig. \ref{SM:2}b, from which the admittance can be calculated as:
\begin{equation}
    S_{11}=\frac{Z-Z_0}{Z+Z_0}
\end{equation}
where $Z_0$ is the impedance of the RF cable which is 50 $\Omega$, and $Z=1/Y$ is the impedance of the device. The admittance obtained from the S$_{11}$ measurement is then fitted using Eq. \ref{eq1} as shown in Fig. \ref{SM:2}b. The fitting parameters are summarized in Table \ref{table1}. The electromechanical coupling efficiency $k_{t,eff}^2$ can be estimated by taking the ratio of capacitance \cite{Blesin:21} $k_{t,eff}^2=C_m/C_0$ which is around 0.2 $\%$ based on the fitting. The mechanical $Q_m$ can be calculated by $1/\omega_mR_mC_m$ which is 270. 

\begin{table*}[t]
    \centering
    \caption{Fitting parameters of MBVD model }
    \begin{tabular}{p{1.6cm}p{1.6cm} p{1.6cm}p{1.2cm}p{1cm}p{1cm}p{1cm}p{2cm} p{1cm} }
 \hline
  $C_0$ & $C_m$ & $L_m$ & $R_m$ & $R_0$ & $R_s$ & $k_{t,eff}^2$ & $\omega_m/2\pi$ & $Q_m$\\
 \hline
2.137 pF & 4.274 fF & 0.677 $\upmu$H & 46.8 $\Omega$ & 45 $\Omega$ & 47 $\Omega$ & 0.2 $\%$ & 2.958 GHz & 270 \\
\hline
\end{tabular}
\label{table1}
\end{table*}

\begin{figure*}[htbp]
\centering
\includegraphics[width=\textwidth]{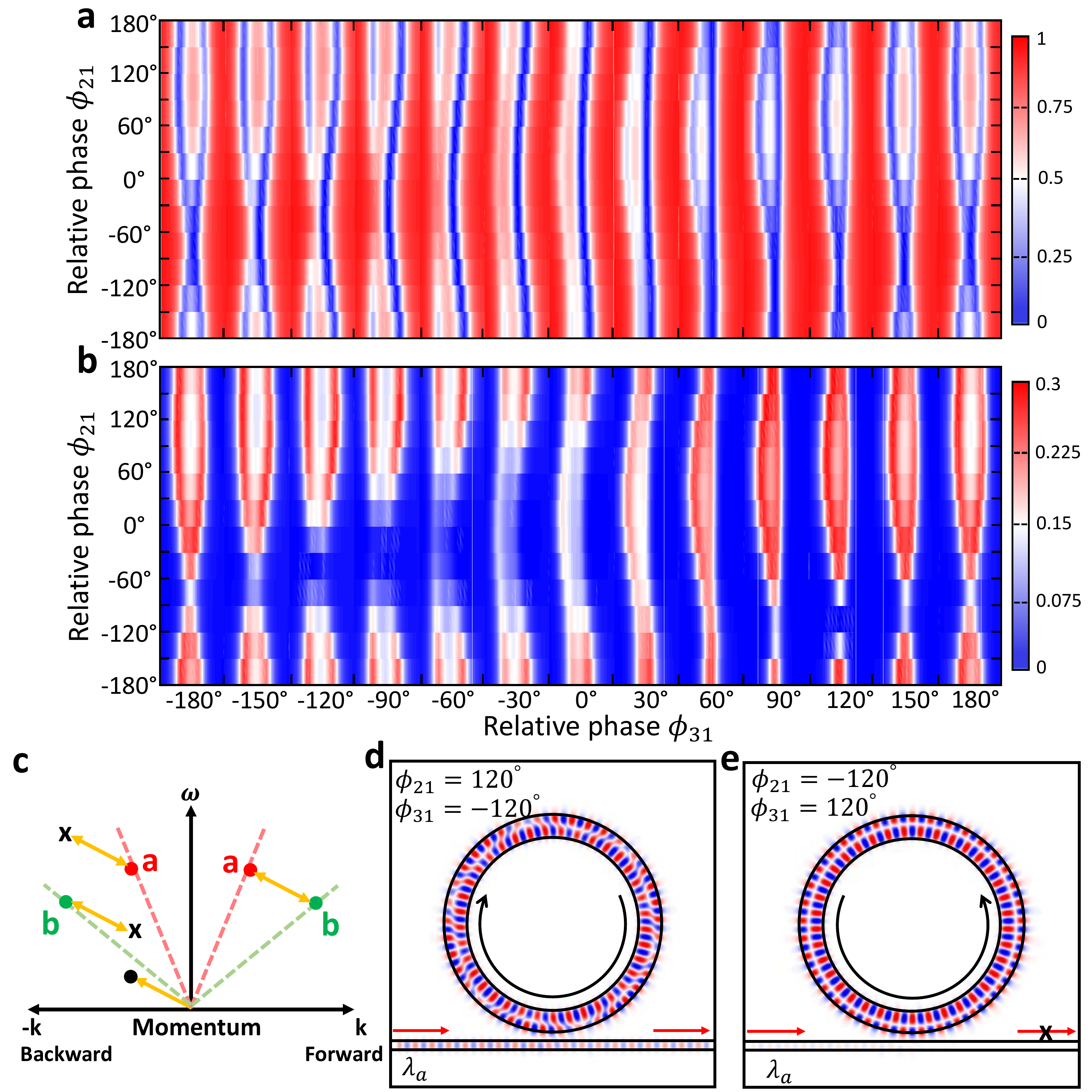}
\caption{\textbf{FDFD numerical simulation of the RF phase dependency}. \textbf{a} Transmission spectra of mode $a$ and \textbf{b} converted sideband in mode $b$ under 2D sweep of the relative phase $\phi_{21}$ and $\phi_{31}$. It shows reasonable matching with the experiment in the main text. Each column is with the spectra span (1.5027 nm, 1.5031nm). \textbf{c} $\omega-k$ space showing the relative position of the two optical modes and the rotational modulation wave. Electric field ($E_z$, out of plane) distribution under reversed phases: \textbf{d} $(\phi_{21}, \phi_{31})=(120^\circ, -120^\circ)$ (perfect phase matching) and \textbf{e} $(\phi_{21}, \phi_{31})=(-120^\circ, 120^\circ)$ (largest phase mismatch). The input light wavelength is at the resonant wavelength of mode $a$ (1502.8 nm). As we change the rotation direction of the modulation wave, the light changes from \textbf{d} transmission to \textbf{e} isolation. 
}
\label{SM:3}
\end{figure*}

\section*{Supplementary Note 4: Numerical simulation of RF phase dependency}
The RF phase dependency is numerically studied in this section which verifies the experiments in the main text. The numerical simulation is conducted through a Finite Difference Frequency Domain (FDFD) algorithm developed by Yu Shi et al. \cite{Shi:16, FDFD}. 
As the optical resonator in the experiment is relatively large (118 $\upmu$m radius) which consumes extremely large simulation time, a much smaller optical ring (3 $\upmu$m radius) is simulated following the same setting as Ref \cite{Shi:18}, where more structure details and optical properties can be found. Despite of that, it shows much similar RF phase dependency behaviour which cross verifies the experiments and numerical model regardless of the size of the optical ring. 

Similar as the experiment, the modulation wave need to counter-propagate with the optical modes to satisfy phase matching (see Fig. \ref{SM:3}c). 
In the simulation, mode $a$ is pumped and mode $b$ is as sideband. Different from previous work \cite{Shi:18}, the phase $(\phi_{21}, \phi_{31})$ are swept in the simulation, and the 2D plots of the spectra of the transmission of mode $a$ and conversion of mode $b$ are as shown in Fig. \ref{SM:3}a,b. By comparing with the experiment in the main text, they show qualitative similarity and will arrive at similar conclusions. The light distribution under reversed phases $(\phi_{21}, \phi_{31})=(120^\circ, -120^\circ)$ (perfect phase matching) and $(\phi_{21}, \phi_{31})=(-120^\circ, 120^\circ)$ (largest phase mismatch) are shown in Fig. \ref{SM:3}d, e with fixed input light at the resonance of mode $a$. Under phase matching (Fig. \ref{SM:3}d), the light transmits through the bus waveguide, and the electric field in the optical ring is a mixture of mode $a$ and $b$ due to the mode coupling. If we reverse the modulation wave's direction (Fig. \ref{SM:3}e), the light is absorbed into the optical ring and the electric field shows the original distribution of mode $a$, indicting no mode coupling is induced. 
\begin{figure*}[htbp]
\centering
\includegraphics[width=\textwidth]{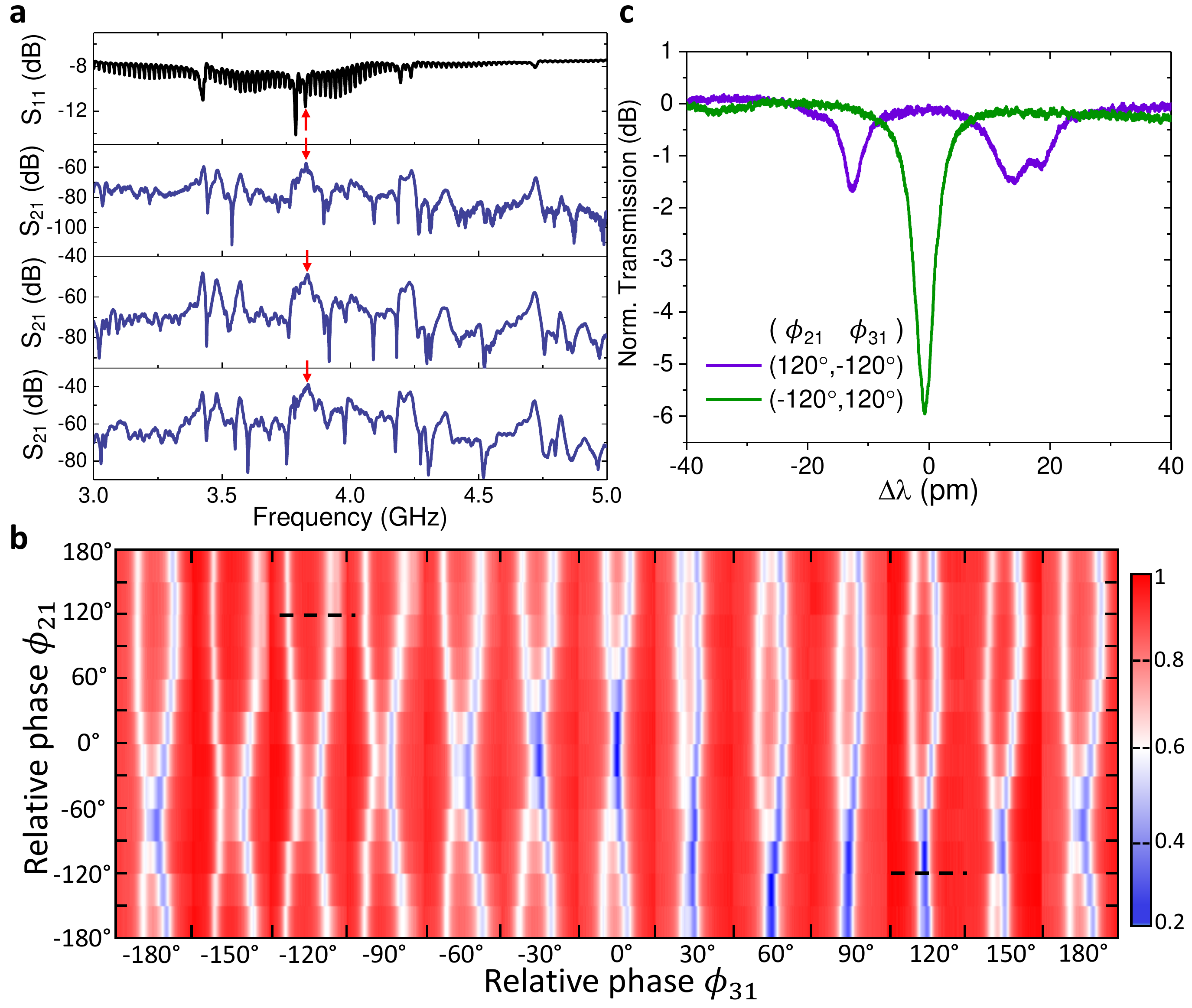}
\caption{\textbf{Measurement of device with 0.1 dB insertion loss}. \textbf{a} From top to bottom are microwave reflection S$_{11}$, optomechanical response S$_{21}$ of the three actuators 1, 2, 3, respectively. Red arrows mark the mechanical mode at 3.833 GHz that is used in the experiments. \textbf{b} Normalized optical transmission of TE$_{00}$ mode under 2D sweep of phases of signals 2 and 3 relative to signal 1, $\phi_{21}$ and $\phi_{31}$. 
Each column is the measured spectrum under the same $\phi _{31}$ with spectral span of (-24 pm, 24 pm) relative to the center wavelength $\lambda_0$ (1553 nm) of TE$_{00}$ mode. \textbf{c} Optical transmission spectra along black dashed lines in \textbf{b} with reversed phases: $(\phi_{21}, \phi_{31})=(120^\circ, -120^\circ)$ (perfect phase matching) and $(\phi_{21}, \phi_{31})=(-120^\circ, 120^\circ)$ (largest phase mismatch).
}
\label{SM:4}
\end{figure*}

\section*{Supplementary Note 5: Measurement of device with smaller insertion loss}
Another device with 950 nm bus-microring gap is measured which shows smaller insertion loss of 0.1 dB. The electro- and opto-mechanical properties are characterized in Fig. \ref{SM:4}a. It shows pretty much similar S$_{11}$ as the device in the main text and the mechanical resonances of the actuators also align with each other. Since the optical mode spacing is around 4 GHz for this device, the HBAR at 3.833 GHz is chosen in the experiment. Similar RF phase dependency measurement is conducted as shown in Fig. \ref{SM:4}b under 20 dBm RF power applied on each actuator. We can make direct comparison with the simulation in Fig. \ref{SM:3}c. This device shows larger mode splitting and maximum of 3.3 GHz is achieved at phase matched case $(\phi_{21}, \phi_{31})=(-120^\circ, 120^\circ)$. Intriguingly, for phase $\phi _{31}$ between $-150^\circ$ and $-60^\circ$, there is always mode splitting regardless of $\phi _{21}$ because of its stronger mode coupling. The optical transmission under reversed phases is shown in Fig. \ref{SM:4}c. Due to the stronger mode coupling $g$ (1.65 GHz), and thus larger optical cooperativity $C$ ($\sim 16$), larger optical transmission at center wavelength can be reached around 98$\%$, corresponding to 0.1 dB insertion loss. However, as this device has larger bus-microring coupling gap, the microring is under-coupled to the bus waveguide which leads to less extinction (6 dB) in the backward direction and thus smaller isolation ratio than the device in the main text. Nevertheless, this measurement tells us that both low insertion loss (0.1 dB) and high isolation ($>20$dB) can be achieved if we design correctly the bus-microring gap in the future.

\section*{Supplementary Note 6: Electromechanical cross-talk between adjacent actuators}

\begin{figure*}[h]
\centering
\includegraphics[width=\textwidth]{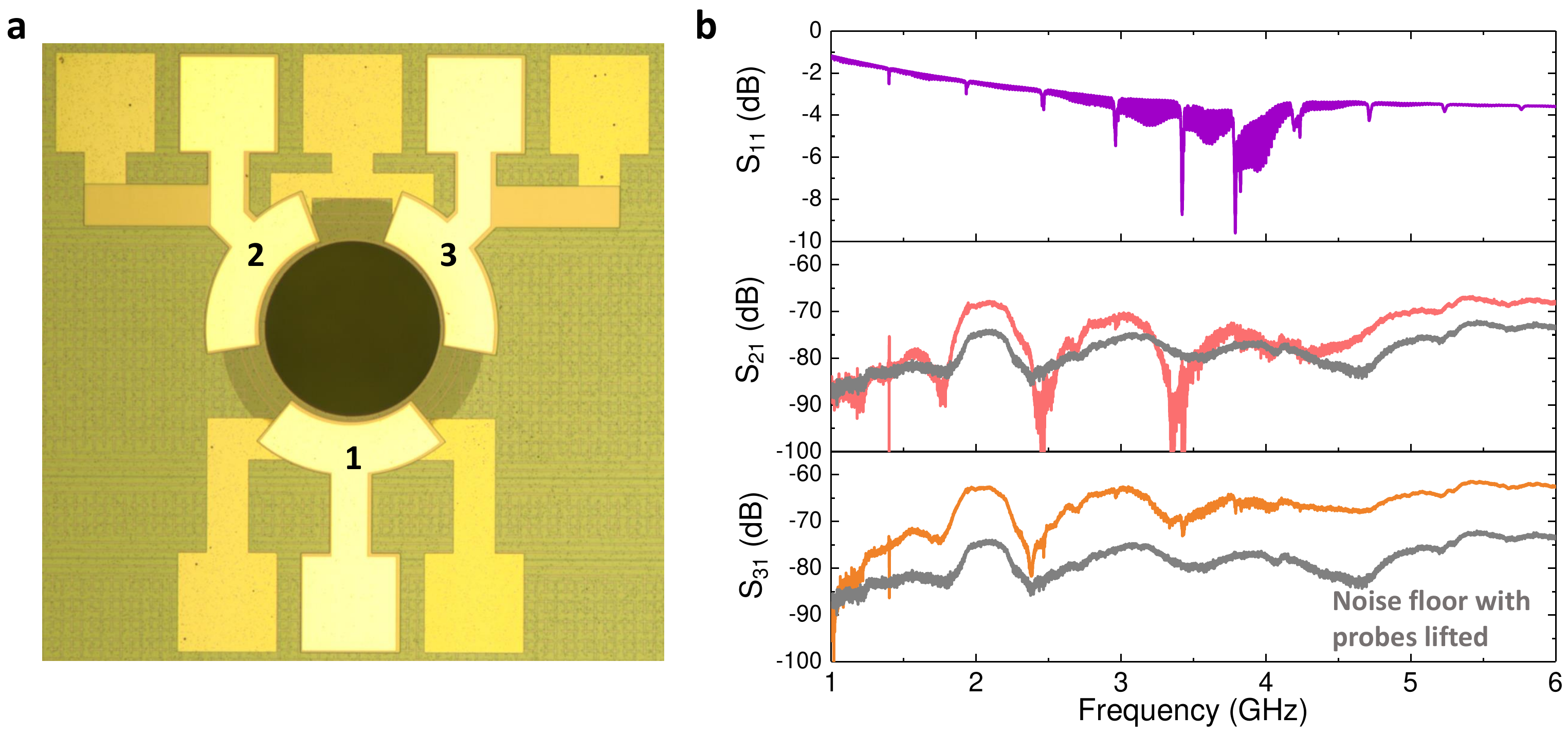}
\caption{\textbf{Electromechanical cross-talk between adjacent actuators}. \textbf{a} Optical microscope image of the device. Three actuators are as labeled. \textbf{b} Electromechanical S$_{11}$ (purple), electrical cross-talk between actuators 2 and 1 S$_{21}$ (pink), and 3 and 1 S$_{31}$ (yellow). The noise floor with probes lifted (gray) is shown for reference. 
}
\label{SM:5}
\end{figure*}

Maintaining small electrical signal cross-talk between adjacent actuators is important for keeping stable relative phases between them. The cross-talk is measured by driving actuator 1 and measuring the output electrical signal from actuator 1 (S$_{11}$), actuator 2 (S$_{21}$), and actuator 3 (S$_{31}$), as shown in Fig. \ref{SM:5}. Due to the rotational symmetry of the device, the cases for driving actuators 2 and 3 are similar. From the results, we can see the cross-talk is well maintained below -60 dB over the measured 1 to 6 GHz range, which means the cross-talk from adjacent actuators is 6 orders of magnitude smaller than the signal applied. The low cross-talk mainly comes from the tight acoustic wave confinement and the center release hole prevents lateral acoustic waves from propagating. The variation of envelope is mainly from the background noise (gray) which is measured by lifting the probes. 

\begin{figure*}[h]
\centering
\includegraphics[width=\textwidth]{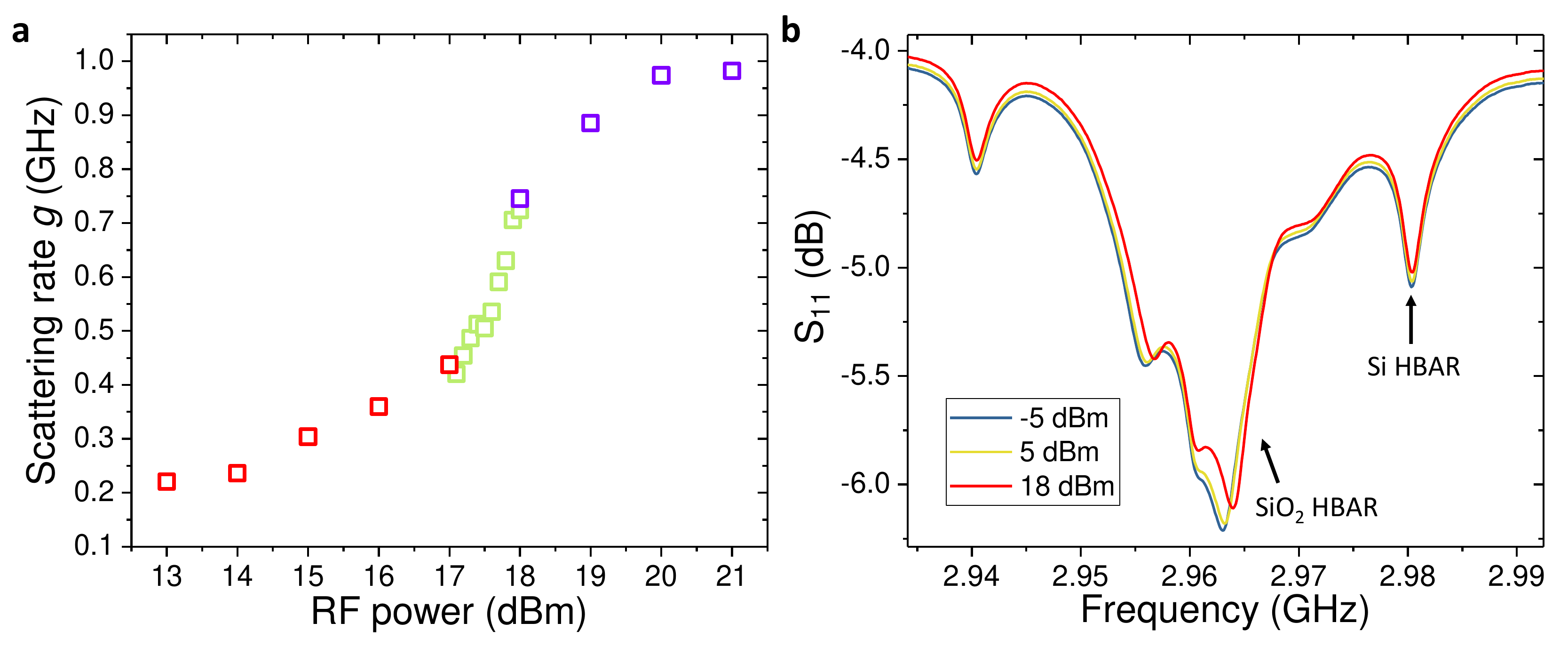}
\caption{\textbf{Thermal heating of RF power}. \textbf{a} Scattering rate $g$ at different RF power. Green squares are fine sweep between 17 and 18 dBm with 0.1 dBm step. Red and purple squares correspond to the same data in the main text. \textbf{b} Electromechanical S$_{11}$ under different RF powers around the HBAR mode used in the main text. The resonances of SiO$_2$ and Si HBAR are as labeled.
}
\label{SM:6}
\end{figure*}

\section*{Supplementary Note 7: Thermal heating effects of high RF power}

\begin{figure*}[htbp]
\centering
\includegraphics[width=\textwidth]{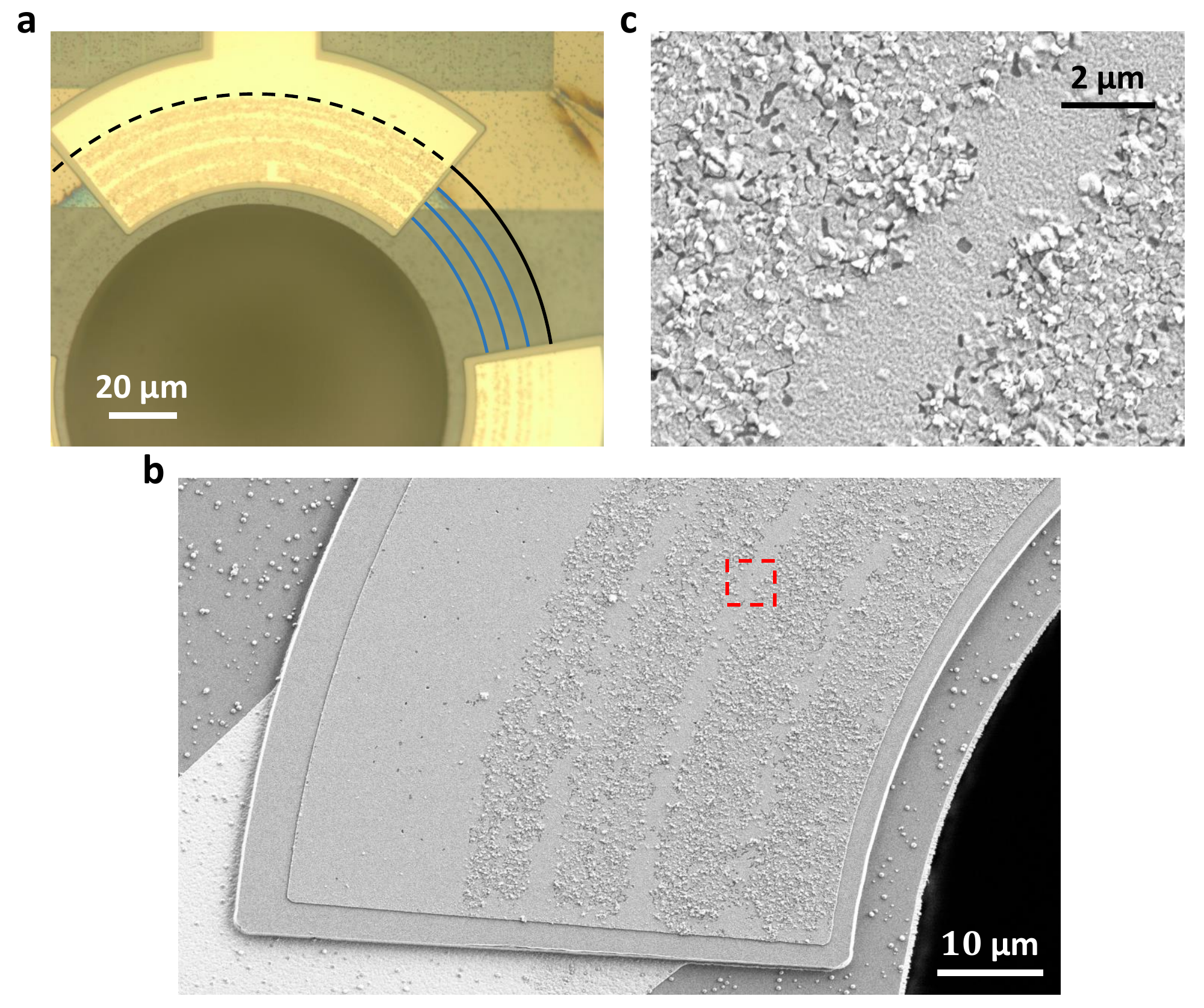}
\caption{\textbf{Degradation of the actuator under 30 dBm RF power}. \textbf{a} Optical microscope image of the device undergoes 30 dBm RF power. Black line denotes the edge of the undercut. Blue lines show the location of Si$_3$N$_4$ microring waveguide and dummy rings. \textbf{b} SEM image showing the corrugated surface of the Al top electrode. \textbf{c} Zoom-in SEM of the region in the red dashed box in \textbf{b}. 
}
\label{SM:7}
\end{figure*}

As discussed in the main text, the thermal heating effect of RF power leads to fast transition from low $g$ at 17 dBm to high $g$ at 18 dBm. Here, detail of how this transition takes place is studied by doing fine sweep between 17 and 18 dBm, as shown in Fig. \ref{SM:6}a. The fine sweep connects the low and high RF power data set. The sweep is conducted back and forth, and no obvious hysteresis is observed. This may be because the sweep speed is slower than the time constant of the thermal heating or it is hidden by the measurement uncertainties. To further explain the origin of the heating effects, electromechanical S$_{11}$ is measured under different RF powers as shown in Fig. \ref{SM:6}b. It can be seen clearly the SiO$_2$ HBAR that is used in the experiments blue shifts as we apply 18 dBm power (red curve) relative to -5 dBm power (blue curve). Intriguingly, due to the better thermal conductance of the Si substrate, the Si HBARs show no observable shifting, which also works as a self-reference to verify that the blue shift of SiO$_2$ HBAR is not from measurement error. The blue shift of the SiO$_2$ HBAR is mainly from the bad thermal conductance of the free-standing SiO$_2$ membrane and the large temperature coefficient of elasticity of SiO$_2$. 

The power handling capability of the piezoelectric actuator is also studied by applying even higher RF powers. For RF power under 27 dBm, the piezoelectric actuator can work continuously for a long time (over several hours) without degradation. However, for RF power bigger than 30 dBm, there is gradual roughing of the top electrode surface, as can be seen in Fig. \ref{SM:7}a. Also, there is burning at the sharp corners of the bottom electrode which is caused by the current crowding under high RF power. Interestingly, the outer edge of the roughing region aligns with the edge of the undercut (black line), which suggests that the degradation is mainly due to the low thermal conductivity of the suspended SiO$_2$ membrane. Also, the degradation is split by three concentric circles into several regions which align with the underneath Si$_3$N$_4$ microring waveguide and dummy ring structures (blue lines) designed for uniform Chemical Mechanical Polishing (CMP). This can be because the existence of these Si$_3$N$_4$ structures scatters the acoustic waves which reduces the energy density around these regions. The roughing can be seen more clearly for the SEM images in Fig. \ref{SM:7}b, c. The top Al layer is corrugated and detached from the bottom AlN. One possible reason for this detachment is the large vertical displacement and vibration under high power loosens the adhesion of Al on AlN. This may explain why the Si$_3$N$_4$ region is not influenced since the acoustic energy is diffused by the scattering. So the corrugation of Al may "undesirably" map out the distribution of acoustic mode.

\begin{figure*}[h]
\centering
\includegraphics[width=14 cm]{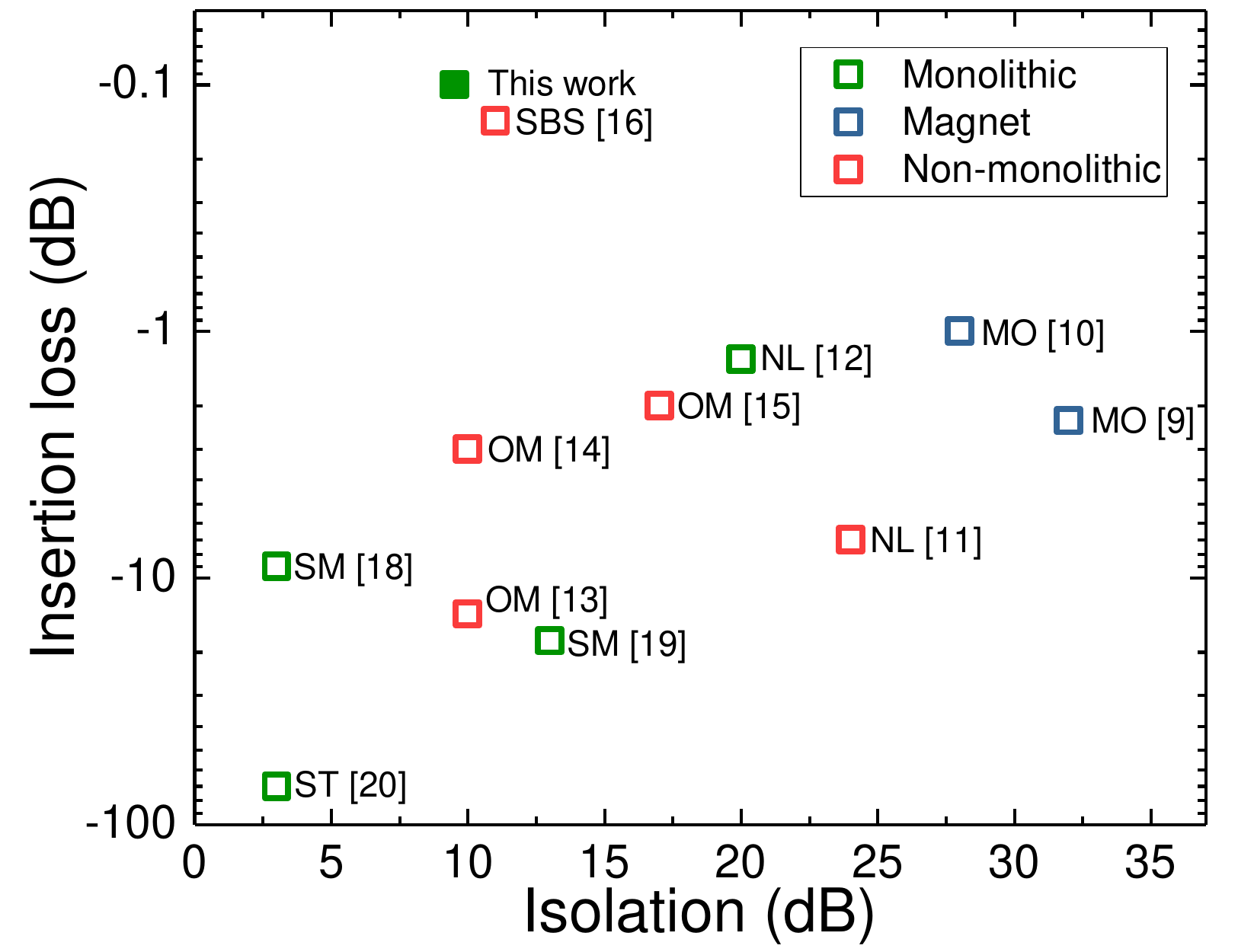}
\caption{\textbf{Comparison of insertion loss and isolation with other experimental realizations}. These works are classified by the level of integration into monolithic (green), and non-monolithic (red), and the works relying on magneto-optical material (blue) is separately listed while others are all magnet-free. We demonstrate the lowest insertion loss among these works. The references are as labeled. 
}
\label{SM:8}
\end{figure*}

\begin{table*}[h]
    \centering
    \caption{Comparison of optical non-reciprocal devices realized in different schemes: MO (magneto-optical), NL (nonlinear optics), OM (optomechanical), SBS (stimulated Brillouin Scattering), SM (Synthetic magnetic field), ST (Spatio-temporal modulation). IL: insertion loss. $a$: Non-reciprocal sideband modulation. $b$: Optical pump power. }
    \begin{tabular}{p{1.1cm}p{1.1cm} p{1.2cm}p{1.87cm}p{1.3cm}p{1.4cm}p{1.7cm}p{1.6cm}p{1.2cm}  }
 \hline
  Year & Scheme & Structure & Material & Isolation & IL (dB) & Bandwidth & Power & CMOS\\
 \hline
2016\cite{Huang:16} & MO & Ring & Si+Ce:YIG & 32 dB & 2.3 & 15 GHz& 10 mW & No\\
2020\cite{Yan:20} & MO & Ring & SiN+Ce:YIG & 28 dB & 1 & 15 GHz&  & No\\
2018\cite{DelBino:18} & NL & Toroid & SiO$_2$ & 24 dB & 7 & 1 MHz& No drive & No\\
2020\cite{YangKY:20} & NL & Ring & Si & 20 dB & 1.3 & 20 GHz& No drive & Yes\\
2016\cite{Ruesink:16} & OM & Toroid &  SiO$_2$ & 10 dB & 14  & 250 kHz & 17 $\upmu$W$^b$ & No\\
2018\cite{Ruesink:18} & OM & Toroid &  SiO$_2$ & 10 dB & 3  & 60 kHz & 60 $\upmu$W$^b$ & No\\
2018\cite{Shen:18} & OM & Sphere &  SiO$_2$ & 17 dB & 2 & 200 kHz & 7.8 mW$^b$ & No\\
2017\cite{Kim:17} & SBS & Sphere &  SiO$_2$ & 11 dB & 0.14 & 400 kHz & 235 $\upmu$W$^b$ & No\\
2014\cite{Tzuang:14} & SM & MZI &  Doped Si & 2.4 dB & & 20 nm& 34 dBm & Yes\\
2021\cite{Kim:21} & SM & Ring &  AlN & 3 dB & 9 & 4 GHz& 16 dBm & Yes\\
2021\cite{Dostart:21} & SM & Ring &  Doped Si & 13 dB & 18 & 2 GHz& -3 dBm & Yes\\
2012\cite{Lira:12} & ST & MZI &  Doped Si & 3 dB & 70 & 200 GHz& 25 dBm & Yes\\
2018\cite{Kittlaus:18} & ST$^a$ & MZI &   Si & 39 dB & NA & 125 GHz& 90 mW$^b$ & Yes\\
2018\cite{Sohn:18} & ST$^a$ & Ring &  AlN & 15 dB & NA& 1 GHz& 18 dBm & Yes\\
2021\cite{Kittlaus:21} & ST$^a$ & MZI &  Si+AlN & 16 dB & NA& 100 GHz& 21 dBm & Yes\\
This work & ST & Ring &  Si$_3$N$_4$+AlN & 10 dB & 0.1-1 & 0.7 GHz& 20 dBm & Yes\\
 \hline
\end{tabular}
\label{table2}
\end{table*}
\section*{Supplementary Note 8: Comparison with current optical non-reciprocal devices}

\noindent This section reviews and summarizes most recent optical non-reciprocal devices realized in different schemes, and compares the performances with our work in Table \ref{table2}. 
The isolation and insertion loss of these works are summarized and plotted in Fig. \ref{SM:8} for better visualization. 
Optical non-reciprocal devices have long been realized using magneto-optical materials which show Faraday effect. However, it is challenging to integrate these materials on chip as they are not compatible with most CMOS processes. 
Nevertheless, there have been exciting advancement towards integrating cerium-substituted yttrium iron garnet (Ce:YIG) on Si photonic chips via wafer bonding \cite{Huang:16}. 
Most recently, Ce:YIG has been grown on SiN by pulsed laser deposition (PLD) \cite{Yan:20}, showing promising performance. However, it still needs bulky external magnet to generate the required external magnetic field to break the Lorentz reciprocity. On the other hand, in the applications for building an optical interface to connect distant superconducting circuits for quantum internet \cite{Awschalom:21}, the applied external magnetic field would unavoidably interfere the operation of superconducting qubits. 
Thus, it is especially desirable for a magnet-free optical isolator in this case.

Optical non-reciprocity has been demonstrated in an optomechanical system through optomechanically induced transparency (OMIT) \cite{Ruesink:16, Shen:16, Ruesink:18, Shen:18}, where an optical pump in one optical mode excites and couples a mechanical resonance with another optical mode. However, most of these demonstrations are realized in a stand-alone optical microresonator (microtoroid or sphere) that supports both optical and mechanical whispering gallery modes (WGMs). This makes it difficult to integrate with PICs in a convenient and reliable way. Moreover, as it requires optical pump, the power of the probe light has to be much smaller than the pump light to maintain stable optomechanical interaction. This limits the dynamic range of the optical power that can be isolated. On the other hand, in an optomechanically induced transparency, the bandwidth of the transparency window is limited by the mechanical linewidth, which limits the isolation bandwidth to below MHz (see Table \ref{table2}). 

From the Table \ref{table2}, most monolithic integrated optical non-reciprocal devices are realized by either optical nonlinearity or dynamical modulation (spatio-temporal modulation and synthetic magnetic field), benefiting from the advanced integration of nonlinear components \cite{Kovach:20} and electro-optical (acousto-optic) modulators \cite{Melikyan:14, WangC:18, HeM:19, Tian:20, Chen:14, Tadesse:14, Shao:19, Tian:21}.

The recent work using the nonlinearity of Si \cite{YangKY:20} achieved large isolation (20 dB) and bandwidth (20 GHz). However, the fact it works for small range of optical power (4 dBm--8 dBm) limits the applicable scenarios. In contrast, dynamic modulation, especially with electrical driving, largely preserves the optical linearity by separating driving and sensing in two different domains. 
The earliest experiments \cite{Lira:12, Tzuang:14} are based on modulating doped Si waveguides in a Mach Zehnder Interferometer (MZI). 
However, the extremely large loss of the doped Si waveguide (70 dB insertion loss) limits the maximum isolation to be 2.4--3 dB. 
In recent years, spatio-temporal modulation using AOM has been implemented in several works \cite{Kittlaus:18, Sohn:18, Kittlaus:21}. 
Limited by either the modulation efficiency or the power handling capability of electrodes, complete mode conversion ($C=1$) hasn't been achieved, and only non-reciprocal sideband modulation is demonstrated. 
Most recently, optical isolators are demonstrated by modulating coupled optical rings through either synthetic Hall effect \cite{Kim:21} or Aharonov-Bohm effect \cite{Dostart:21}, but with limited isolation \cite{Kim:21} and large insertion loss \cite{Dostart:21}. 
In comparison, our HBAR resonator shows larger power handling capability because of its wide electrode area (and thus small resistance). Also, the tight confinement of acoustic energy in thin oxide membrane largely improves the modulation efficiency (by 100 times) compared with previous unreleased Si HBAR \cite{Tian:20}. With these features, we demonstrated the lowest insertion loss among these works, and comparable isolation under decent amount of RF power applied.

\spacing{1}

\section*{References}
\bigskip
\bibliographystyle{naturemag}
\bibliography{bibliography}